\UseRawInputEncoding

\documentclass[pra,aps,twocolumn,epsfig]{revtex4-2}

\usepackage{graphicx}
\usepackage{dcolumn}
\usepackage{bm}

\begin{document}


\title{Reduction of thermodynamic uncertainty by a virtual qubit}


\author{Yang Li}
\affiliation{Department of Physics, School of Science, Tianjin University, Tianjin 300072, China}

\author{Fu-Lin Zhang}
\email[Corresponding author: ]{flzhang@tju.edu.cn}
\affiliation{Department of Physics, School of Science, Tianjin University, Tianjin 300072, China}

\date{\today}

\begin{abstract}
The thermodynamic uncertainty relation (TUR) imposes a fundamental constraint between current fluctuations and entropy production, providing a refined formulation of the second law for micro- and nanoscale systems. Quantum violations of the classical TUR reveal genuinely quantum thermodynamic effects, which are essential for improving performance and enabling optimization in quantum technologies.
In this work, we analyze the TUR in a class of paradigmatic quantum thermal-machine models whose operation is enabled by coherent coupling between two energy levels forming a \emph{virtual qubit}. Steady-state coherences are confined to this virtual-qubit subspace, while in the absence of coherent coupling the system satisfies detailed balance with the thermal reservoirs and supports no steady-state heat currents.
We show that the steady-state currents and entropy production can be fully reproduced by an effective classical Markov process, whereas current fluctuations acquire an additional purely quantum correction originating from coherence. As a result, the thermodynamic uncertainty naturally decomposes into a classical (diagonal) contribution and a coherent contribution. The latter becomes negative under resonant conditions and reaches its minimum at the coupling strength that maximizes steady-state coherence. 
We further identify the conditions for minimizing the thermodynamic uncertainty, and the criteria for surpassing the classical TUR bound in the vicinity of the reversible limit.
%
 
 \end{abstract}

\maketitle

\section{Introduction}

Thermal machines provide paradigmatic platforms for studying the fundamental principles
governing the conversion between work, heat, and information.
In the context of quantum thermodynamics, they continue to play a central role, where
genuinely quantum effects participate in and modify these conversion processes 
\cite{ARPC2014quantum,
Goold_2016,
cangemi2024quantum}.
 At the microscopic scale, both thermal and quantum
fluctuations strongly influence the stability and precision of thermodynamic functionalities,
rendering a quantitative understanding of fluctuations an essential component of
quantum thermodynamics.

A central result addressing this issue is the thermodynamic uncertainty relation (TUR),
originally discovered by Barato and Seifert \cite{Barato2015Thermodynamic} in the context of biomolecular processes.
It can be expressed as
\begin{equation}\label{TUR}
\mathcal{Q}
\equiv
\frac{\sigma\, \mathrm{Var}(J)}{ J^2}
\ge 2 ,
\end{equation}
where $J$ denotes the steady-state heat current flowing between the system and one of
its thermal reservoirs, $\mathrm{Var}(J)$ its variance in the long-time limit, and
$\sigma$ the corresponding entropy production rate.
This inequality implies that suppressing current fluctuations,
quantified by $\mathrm{Var}(J)/J$, necessarily comes at the cost of
increased thermodynamic dissipation, characterized by the normalized dissipation rate
$\sigma/J$.
Since its original formulation, 
the TUR,
together with the associated fluctuation theory, has been applied and extended
in a wide range of settings, including thermal transport~\cite{Saryal2019},
stochastic clocks~\cite{Barato2016}, quantum batteries~\cite{sarkar2025fluctuation},
as well as finite-time~\cite{Pietzonka2017,Horowitz2017,Otsubo2020} and periodically
driven systems~\cite{Koyuk2020}.

Recently, the role of quantum coherence in TURs, including its impact on
violations of the classical TUR, has been investigated in a variety of quantum
systems
\cite{Kalaee2021,bayona2021thermodynamics,JPA2021TUR,PRA2023TUR,PRA2025,mohan2025coherent,mohan2025enhancing},
and several quantum extensions and formulations of TURs have been proposed
\cite{Guarnieri2019,
Hasegawa2020,
Hasegawa2021,
VanVu2022,
Prech2023,
Salazar2024,
LiuGu2025}.
In this work, we adopt the original definition of the TUR introduced by
Barato and Seifert and investigate how quantum features reduce the uncertainty
product and enable violations of the classical bound.


We present a unified analysis of the thermodynamic uncertainty for a class of thermal machines whose operation is enabled by coherent coupling between two energy levels forming a virtual qubit \cite{PRE2012virtual}.
This coupling may be either time-independent or periodically driven, and quantum coherence in the steady state is confined to this virtual-qubit subspace.
In the absence of the coherent coupling, the steady state of the system in contact with multiple thermal reservoirs obeys detailed balance and supports no steady-state heat currents.

The motivation for studying this class of thermal machines is as follows.
For time-independent coupling, it includes paradigmatic models of autonomous thermal machines 
\cite{ARPC2014quantum,
Goold_2016,
cangemi2024quantum},
such as the three-qubit absorption refrigerator
\cite{PRL2010small,JPA2011smallest},
whose study has stimulated significant interest in autonomous thermal machines.
For time-dependent coupling, this class encompasses driven quantum machines,
including the driven qubit and the Scovil--Schulz-DuBois three-level maser 
\cite{Scovil1959,
Geva1994,
klatzow2019experimental,
Kalaee2021,
bayona2021thermodynamics}
(referred to as a driven qutrit in this work),
which were among the earliest models shown to exhibit coherence-induced violations
of the classical TUR bound.
%
In both autonomous and driven settings, the steady-state heat currents and the associated entropy production can be described by classical Markovian stochastic processes, as has been noted in specific models \cite{Gonz_lez_2017,Kalaee2021}. 
As a consequence, genuinely quantum effects enter exclusively through the second-order
current fluctuations.
Owing to the confinement of the coherent coupling and the resulting steady-state coherence to the virtual-qubit subspace, this minimal steady-state structure allows us to clearly isolate the contribution of quantum coherence to current fluctuations and, consequently, to the thermodynamic uncertainty.


In this work, the coupling to thermal reservoirs is modeled by Lindblad-type
dissipators, and the combined effects of the coherent coupling within the virtual
qubit and the system--reservoir interactions are treated within a \emph{local approach} \cite{EPL2014local,ACMP2015nonequilibrium,
EPL2016perturbative,Wang2023}.
Within this framework, the coherent coupling does not modify the structure of the
dissipators, which corresponds to assuming that the coherent coupling strength is
comparable to the system--reservoir coupling, as reflected by rates $p_c$ and $p_I$
(introduced in Sec.~\ref{SecTUR}) of the same order.
The distinction between the local approach and the\emph{ global approach} lies in the
choice of representation employed in the perturbative treatment.
The local approach is formulated in the eigenbasis of the bare system Hamiltonian,
whereas the global approach is defined in the eigenbasis of the full Hamiltonian,
including the coherent coupling \cite{Wang2023}.
Consequently, physical notions such as energy and coherence must be interpreted
within the representation associated with the chosen approach.
For time-independent coupling, the steady state obtained within the global
approach is a statistical mixture of eigenstates of the full Hamiltonian and thus
exhibits no steady-state coherence.
By contrast, for the driven model under resonant conditions—where violations of the classical TUR were reported in Ref.~\cite{bayona2021thermodynamics}—the resulting master equation is formally equivalent to that obtained within the local approach, as shown in  Appendix \ref{global}.

For the class of models considered in the present work, we develop a systematic procedure to evaluate the thermodynamic uncertainty.
The central step consists in solving three diagonal operators determined by the dissipators, which satisfy a closed set of linear equations.
We show that the steady-state currents and the associated entropy production can always be exactly reproduced by an equivalent classical Markov process, whereas current fluctuations acquire an additional purely quantum contribution originating from steady-state coherence.
As a consequence, the thermodynamic uncertainty naturally decomposes into a diagonal (classical) part and a coherent part. The former obeys the classical thermodynamic uncertainty bound, while the latter is strictly negative under resonant conditions.
Remarkably, the coupling strength that minimizes the coherent contribution (i.e., maximizes its absolute value) simultaneously maximizes the steady-state coherence.
We further identify the optimization conditions for the thermodynamic uncertainty and the criteria for surpassing the classical bound in the vicinity of the reversible limit.

This paper is organized as follows.
In Sec. \ref{SecModel}, we introduce the class of models considered in this work and the corresponding master equation.
In Sec. \ref{SecTUR}, we present a unified analysis of the TUR for this class of models. We establish the equivalence between driven and time-independently coupled machines, clarify the relations between different heat currents, and construct the effective classical models corresponding to the underlying quantum dynamics. Based on this analysis, we provide a systematic procedure to evaluate the thermodynamic uncertainty, derive formal expressions, and perform the corresponding optimization analysis.
In Sec. \ref{SecExamp}, we illustrate the general framework using representative physical examples, including a driven qubit and a two-qubit heat-transport model.
Finally, in Sec. \ref{SecConcl}, we summarize our findings and discuss possible extensions, with technical details and additional examples presented in the Appendices.

\section{Model} \label{SecModel}

We consider a class of finite-dimensional quantum thermal machines characterized by the bare Hamiltonian
\begin{eqnarray}
H_0 = \sum_k E_k |\Phi_k\rangle \langle \Phi_k|,
\end{eqnarray}
with $|\Phi_k\rangle$ and $E_k$ denoting its eigenstates and eigenvalues, respectively.
Throughout this work, concepts such as populations, coherences, and diagonal or off-diagonal elements are defined with respect to the eigenbasis $\{|\Phi_k\rangle\}$.
The states $|\Phi_0\rangle$ and $|\Phi_1\rangle$ form a \textit{virtual qubit}, 
which is coherently coupled via the interaction Hamiltonian
\begin{eqnarray}\label{HIt}
H_I = g \left( |\Phi_0\rangle \langle \Phi_1| e^{-i \omega_d t} + |\Phi_1\rangle \langle \Phi_0| e^{i \omega_d t} \right),
\end{eqnarray}
where $g$ is the coupling strength. A nonzero driving frequency $\omega_d \neq 0$ corresponds to a time-dependent external drive, while $\omega_d = 0$ represents a stationary coupling.
The total Hamiltonian of the thermal machine is then
\begin{eqnarray}
H_s = H_0 + H_I.
\end{eqnarray}
In this work, we treat both time-dependent driving and stationary coupling in a unified manner.

\begin{figure}
 \includegraphics[width=8.1cm]{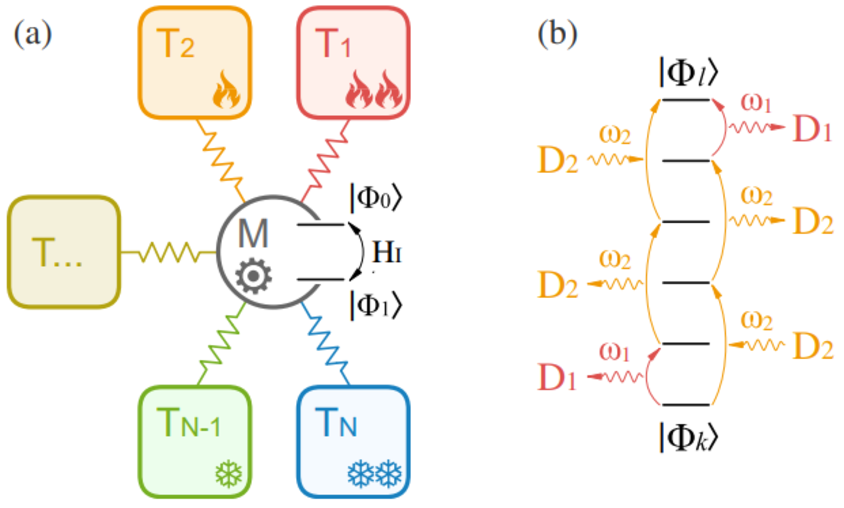}
 \caption{
(Color online)
 Scheme of our device. 
    (a) The thermal machine $M$ is in contact with multiple reservoirs at temperatures 
    $T_1, \ldots, T_N$. 
    Two of its states, $|\Phi_0\rangle$ and $|\Phi_1\rangle$, are coherently coupled through the interaction Hamiltonian $H_I$.
    When $H_I$ vanishes, the system reaches a detailed-balance steady state, i.e., there is no net transition between any pair of states.
    The detailed-balance condition requires that, when multiple transition paths exist between two states,
    the algebraic sum of the numbers of photons with frequency $\omega_i$ emitted to reservoir $i$
    (with absorption counted as negative) is identical for all such paths.
    (b) Example of two transition paths from $|\Phi_k\rangle$ to $|\Phi_l\rangle$.
    Along both paths, the number of photons with frequency $\omega_1$ emitted to the reservoir characterized by the dissipator $D_1$ is $1$,
    and the number of photons with frequency $\omega_2$ emitted to the reservoir associated with $D_2$ is $0$.
    }\label{modelfig}
\end{figure}

 As shown in Fig.~\ref{modelfig}, the system $M$  is coupled to several thermal reservoirs at temperatures 
    $T_1, \ldots, T_N$.  
 These thermal contacts, together with the coherent coupling $H_I$, drive the system to a steady state in which directed transitions occur between its internal states, giving rise to stable heat currents between the system and the reservoirs.
Such heat currents enable the machine to perform thermodynamic tasks, including work extraction and refrigeration.
In the absence of $H_I$, the steady state 
\begin{eqnarray}
\tau = \sum_k P_k\, |\Phi_k\rangle\langle\Phi_k|,
\end{eqnarray}
is diagonal in the eigenbasis $\{|\Phi_k\rangle\}$ and obeys detailed balance with each coupled reservoir, implying the absence of net transitions between any pair of states and vanishing heat currents between the system and the reservoirs.

The dissipative effect of each reservoir on the system is assumed to be described by a Lindblad-type dissipator\cite{Lindblad1976,BookOpenSys}, which drives transitions and decoherence within the eigenbasis $\{|\Phi_k\rangle\}$ of $H_0$, while $H_I$ does not affect these dissipators; that is, the system obeys a master equation based on the local approach\cite{Wang2023}.
Without loss of generality, one can assume that each reservoir is coupled to the system $M$ via a single mode, with frequency $\omega_i$  for the $i$-th reservoir. A reservoir with multiple frequencies interacting with the system can then be regarded as multiple single-mode reservoirs at the same temperature.
As a result, the system evolves according to the master equation (we use $\hbar = k_B = 1$ in the following) \cite{Lindblad1976}
\begin{eqnarray}\label{MasterEq}
    \dot{\rho}= -i [H_s,\rho]+\mathcal{D}(\rho),
\end{eqnarray}
where $\mathcal{D} = \sum_{i} D_i $ and the dissipator takes the form
\begin{eqnarray}\label{Di}
    D_i (\rho)&=& p_i R_i \left(\Gamma_i^{\dagger} \rho \Gamma_i -\frac{1}{2}\{\Gamma_i \Gamma_i^{\dagger}, \rho  \}\right) \nonumber\\
    &\ & +p_i \bar{R}_i \left(\Gamma_i \rho \Gamma_i^{\dagger} -\frac{1}{2}\{\Gamma_i^{\dagger} \Gamma_i, \rho  \}\right).
\end{eqnarray}
Here, $p_i>0$ denotes the strength of the reservoir's influence on the system, with the thermal occupation factors $R_i + \bar{R}_i = 1$ and $R_i / \bar{R}_i = \exp(-\omega_i / T_i)$. 
 The transition operator $\Gamma_i$ is a linear combination of $|\Phi_k\rangle \langle \Phi_l|$ satisfying $E_l - E_k = \omega_i$.

We assume that any two eigenstates $|\Phi_k\rangle$ and $|\Phi_l\rangle$ are connected by at least one transition path generated by the dissipators in Eq.~(\ref{MasterEq}). 
Otherwise, the system can be regarded as a collection of two or more thermodynamically isolated systems, 
each consisting of states that are internally connected by dissipative transition paths.
This assumption guarantees the uniqueness of the steady state $\tau$ in the absence of  $H_I$.
Under this condition,  the detailed-balance condition requires the steady-state populations of $|\Phi_k\rangle$ and $|\Phi_l\rangle$ to satisfy
$
    P_k / P_l = \exp\!\left[\sum_i \left(- n_i \omega_i / T_i\right)\right].
$
Here, $n_i$ denotes the number of photons with frequency $\omega_i$ emitted into the $i$th reservoir; absorption corresponds to negative $n_i$.
If the transition path is not unique, the values of $n_i$ must be identical for all allowed paths, ensuring that detailed balance is not violated by changes in the reservoir temperatures.
Figure~\ref{modelfig}(b) illustrates an example with two distinct transition paths.
The coherent coupling drives the system out of detailed balance and thereby establishes a steady heat current between the system and the reservoirs.

When the right-hand side of the master equation~(\ref{MasterEq}) 
(i.e., the Liouvillian)
 acts on a diagonal state, only the interaction Hamiltonian $H_I$ generates off-diagonal elements, and these coherences remain confined to the virtual-qubit subspace. Accordingly, transition processes of the form
$a|\Phi_l\rangle\langle\Phi_k| + b|\Phi_m\rangle\langle\Phi_k|$
can be excluded from the structure of the jump operators $\Gamma_i$ and $\Gamma_i^\dagger$.
Furthermore, we require that the dissipators do not generate coherences
outside the virtual-qubit subspace when acting on
$|\Phi_0\rangle\langle\Phi_1|$ or $|\Phi_1\rangle\langle\Phi_0|$.
This condition constrains the admissible forms of these jump operators,
which are analyzed in detail in Appendix~\ref{Gamma01}.

\section{THERMODYNAMIC UNCERTAINTY RELATION} \label{SecTUR}

Based on the general structure shared by the class of models, 
we now present a unified analysis of the thermodynamic uncertainty.
While the present section focuses on developing the general framework and formal results, 
the resulting systematic procedure for evaluating the thermodynamic uncertainty will be summarized at the beginning of next setion.

\textit{Removing time dependence.}
Our time-dependent driven model can always be mapped onto an equivalent stationary coupling system by choosing an appropriate rotating reference frame. Crucially, the photon number currents exchanged between the system and the reservoirs in the stationary model are exactly identical to those of the original driven system.
Specifically, introducing the rotating transformation
\begin{eqnarray}\label{Rszw}
\mathcal{R}_s  = \exp(- i G t),
\end{eqnarray}
the density matrix in the rotating frame is defined as
\(\rho^{\rm rot} = \mathcal{R}_s \, \rho \, \mathcal{R}_s^{\dagger} .\)
Its time evolution obeys the master equation
\begin{eqnarray}\label{MasterEqRot}
\dot{\rho}^{\rm rot} 
= - i \big[ H_s^{\rm rot}, \rho^{\rm rot}  \big]
+ \mathcal{D}\!\left( \rho^{\rm rot}  \right),
\end{eqnarray}
where the effective Hamiltonian in the rotating frame is given by
$
H_s^{\rm rot} = H_s\big|_{t=0} + G ,
$
which is time independent.
Here, the generator $G$ is diagonal in the eigenbasis $\{|\Phi_k\rangle\}$, and its eigenvalues are chosen such that the explicit time dependence of the Hamiltonian is removed while the dissipators remain unchanged. 
In the Appendix \ref{NoTime}, we employ a microscopic system--reservoir model to elucidate the construction and physical meaning of the operator $G$, and to demonstrate that the steady-state currents generated by Eq.~(\ref{MasterEqRot}) coincide with those of the driven system.

Among the diagonal terms of $H_s^{\rm rot}$, the only contribution relevant for the steady state and the photon number currents is the detuning parameter
\begin{eqnarray}
\Delta = E_0 - E_1 - \omega_d .
\end{eqnarray}
For a genuinely stationary coupling model, thermodynamic consistency of the local master equation requires $E_0 - E_1=\omega_d = 0$~\cite{Wang2023}.
In summary, \textit{in the following we only need to consider the master equation
Eq.~(\ref{MasterEq}) in the stationary-coupling case}, with
$\Delta = E_0 - E_1$ understood as the detuning between the system energy-level
splitting and the driving frequency.
%

\textit{Relations among currents.}
We restrict our attention to the TUR expressed in terms of the heat current and its variance for those reservoirs with photon numbers $n_i \neq 0$ along the transition paths connecting $|\Phi_0\rangle$ and $|\Phi_1\rangle$.
Reservoirs that do not participate in these transition paths, or that exchange a net number of photons equal to zero along a given channel, have vanishing current cumulants of all orders in the long-time limit.
Furthermore, the currents associated with the two reservoirs carrying nonzero photon numbers along the same transition path are fully correlated.
As a consequence, one may choose either of them to formulate the TUR and obtain identical results.
This can be understood intuitively as follows: during one elementary cycle completed by the system through the transition channels and the interaction Hamiltonian $H_I$, the numbers of photons absorbed or emitted along the two channels are strictly proportional.
A rigorous proof of these statements is provided in the Appendix \ref{G0G12}.

\textit{Classical counterpart and currents.}
The photon-number current flowing from the system into the \(i\)-th reservoir can be obtained
by evaluating the action of the corresponding jump operators on the density matrix and
taking the trace. Explicitly, one finds
\begin{eqnarray}\label{Ji}
  J_i &=& p_i \, \mathrm{Tr}\!\left(-
  R_i\, \Gamma_i^{\dagger}\, \rho \, \Gamma_i
  + \bar{R}_i\, \Gamma_i\, \rho \, \Gamma_i^{\dagger}
  \right) \nonumber\\
      &= &   p_i \, \mathrm{Tr}\!\left(-
  R_i\, \Gamma_i^{\dagger}\, \rho_d \, \Gamma_i
  + \bar{R}_i\, \Gamma_i\, \rho_d \, \Gamma_i^{\dagger}
  \right),
\end{eqnarray}
where \(\rho_d\) denotes the diagonal (classical) part of   \(\rho\).
Moreover, one may construct a "classical" counterpart of the system described by the
master equation~(\ref{MasterEq}) with a stationary coupling.
In this classical description, the coherent interaction \(H_I\) 
is replaced by an effective dissipator \(D_c\), and the dynamics is
governed by the master equation
\begin{equation}\label{MasterEqClassical}
  \dot{\rho}
  = -i [H_0,\rho]+\mathcal{D} (\rho ) + D_c(\rho ),
\end{equation}
where the dissipator \(D_c\) takes the same Lindblad form as in Eq.~(\ref{Di}), with a
strength parameter
\begin{equation}\label{pc}
  p_c = \frac{4 g^2 \gamma}{\gamma^2 + \Delta^2},
\end{equation}
a jump operator
$ 
 \Gamma_c = |\Phi_1\rangle \langle \Phi_0|,
$
and a rate \(R_c = 1/2\).
Here \(\gamma\) denotes the decoherence rate, defined through
 $ 
\mathcal{D} \!\left(|\Phi_1\rangle \langle \Phi_0|\right)
  = - \gamma \, |\Phi_1\rangle \langle \Phi_0| .
$
For a diagonal density matrix, the evolution governed by
Eq.~(\ref{MasterEqClassical}) is fully equivalent to a classical Markov process
for the populations.
Moreover, the steady state of Eq.~(\ref{MasterEqClassical}) is itself diagonal and
coincides with the diagonal part of the steady state obtained from
Eq.~(\ref{MasterEq}) [see Appendix \ref{App_Classical} and \ref{FCS}], while generating exactly the same steady-state currents \(J_i\)

One may further define, in analogy with Eq.~(\ref{Ji}), a current \(J_c\) associated with
the dissipator \(D_c\) in the classical counterpart.
In the steady state, \(J_c\) is equal to the net number of transitions from
\(|\Phi_0\rangle\) to \(|\Phi_1\rangle\) induced by the interaction \(H_I\) in the original
model.
Furthermore, it satisfies the relation
\begin{equation}\label{Ji}
  J_i = - n_i J_c ,
\end{equation}
with the currents flowing into the other reservoirs.
This relation can be established by following a line of reasoning analogous to that presented in Appendix~\ref{G0G12}.
Once the diagonal density matrix
\(\rho_d = \sum_k q_k\, |\Phi_k\rangle\langle\Phi_k|\) is determined, the current \(J_c\)
admits the simple expression
\begin{equation}\label{Jc}
  J_c = \frac{p_c}{2}\, r ,
\end{equation}
where \(r = q_0 - q_1\).
Furthermore, the  dissipator $D_c$ can be interpreted as coupling the system
to a reservoir at infinite temperature.
As a consequence, the entropy production rates associated with the steady-state heat
currents of both the original system described by Eq.~(\ref{MasterEq}) and its classical
counterpart governed by Eq.~(\ref{MasterEqClassical}) can be expressed in an identical form,
namely
\begin{equation}
  \sigma
  = \sum_i \frac{J_i \, \omega_i}{T_i}
 = J_c \ln\!  \frac{P_0}{P_1}  
  = \frac{p_c}{2}\, r \,
    \ln\! \frac{P_0}{P_1}  .
\end{equation}
This quantity is always non-negative, since$r$ and
$P_0 - P_1$  share the same sign \cite{PRE2012virtual}. 
These results demonstrate that the steady-state heat currents and entropy production
in our model can always be faithfully reproduced by an effective classical Markov
process.

\textit{Variance of the current and thermodynamic uncertainty.}
We now evaluate the variance $\mathrm{Var}(J_i)$ of the current $J_i$ 
and the thermodynamic uncertainty
using the counting-field approach \cite{JPA1999general,esposito2009nonequilibrium,bruderer2014inverse}. 
The detailed derivation
is provided in Appendix~\ref{FCS}.
Within this framework, a counting-field parameter is introduced into the
dissipator $D_i$ of the master equation~(\ref{MasterEq}).
Treating the modified master equation perturbatively, one finds that
$\mathrm{Var}(J_i)$ is fully determined by the jump operators associated with the
$i$th reservoir and the steady state of the system.
As a consequence, the current variance naturally decomposes into two distinct
contributions: a classical part originating from the diagonal component of the
steady state, $\rho_d$, and a genuinely quantum part arising from the coherent
(off-diagonal) component, $\rho_c$,
\begin{equation}
\mathrm{Var}(J_i)
=
\mathrm{Var}_d(J_i)
+
\mathrm{Var}_c(J_i).
\end{equation}
The classical part $\mathrm{Var}_d(J_i)$ coincides exactly with the current 
variance obtained from the classical counterpart described by 
Eq.~(\ref{MasterEqClassical}),
\begin{equation}
\mathrm{Var}_d(J_i) = \frac{1}{2} p_c q n_i^2
- 2 J_i^2 \, \mathrm{Tr}\,\xi_d ,
\end{equation}
where $q=q_0 + q_1$ and the diagonal operator $\xi_d$ satisfies
\begin{eqnarray} \label{xidzw}
&\langle \Phi_0 | \xi_d | \Phi_0 \rangle =  \langle \Phi_1 | \xi_d | \Phi_1 \rangle ,& \nonumber \\
&\mathcal{D}(\xi_d)
= \rho_d - \frac{1}{r}\left(q_0 |\Phi_1\rangle \langle \Phi_1| - q_1 |\Phi_0\rangle \langle \Phi_0| \right). &
\end{eqnarray}  
The coherent contribution is given by
\begin{equation}
\mathrm{Var}_c(J_i)
= -2 \frac{1}{\gamma} \frac{\gamma^2 - \Delta^2}{\gamma^2 + \Delta^2} \frac{r}{r_0} J_i^2 .
\end{equation}

Combining the entropy production rate, the current, and its variance derived
above, 
we obtain the thermodynamic uncertainty $\mathcal{Q}$ defined in
Eq.~(\ref{TUR}) in the explicit form
\begin{equation}
\mathcal{Q}\label{QdQc}
=
\underbrace{
\ln \frac{P_0}{P_1}
\bigg(
\frac{q}{r}
-
p_c\, r\, \mathrm{Tr}\,\xi_d
\bigg)
}_{\equiv\,\mathcal{Q}_d}
\underbrace{
- \ln \frac{P_0}{P_1}
\frac{p_c\, r^2}{r_0 \gamma}
\frac{\gamma^2 - \Delta^2}{\gamma^2 + \Delta^2}
}_{\equiv\,\mathcal{Q}_c}.
\end{equation}
In direct correspondence with the decomposition of the current variance, the
thermodynamic uncertainty  naturally separates into a diagonal
(classical) contribution $\mathcal{Q}_d$ and a coherent contribution
$\mathcal{Q}_c$.
The diagonal part $\mathcal{Q}_d$ coincides with the result of the classical
Markovian dynamics described by Eq.~(\ref{MasterEqClassical}) and therefore
satisfies the standard TUR
$\mathcal{Q}_d \ge 2$.
By contrast, when the detuning $\Delta$ is smaller than the linewidth $\gamma$,
the coherent part $\mathcal{Q}_c$ is negative, indicating a 
quantum advantage that reduces the thermodynamic uncertainty.
This also implies that any violation of the classical bound in our model,
namely $\mathcal{Q}<2$, can be unambiguously attributed to the coherent
contribution $\mathcal{Q}_c$, or equivalently to the coherent part of
the variance $\mathrm{Var}_c(J_i)$.

\textit{Optimization of the thermodynamic uncertainty. }
We now consider the optimization of the thermodynamic uncertainty $\mathcal{Q}$ for the present model
and identify the conditions under which the classical bound $\mathcal{Q} \ge 2$ is violated.
The detailed derivations are presented in Appendices~\ref{Qcmin}, \ref{Qmin}, and \ref{r00}.
The system dynamics are determined by the interplay between the dissipators $\{D_i\}$ and the
Hamiltonian $H_0 + H_I$, and thus depend on the parameters $\{p_i, R_i, g, \omega_d, E_k\}$.
In the thermodynamic uncertainty, however, the dependence on $\{\omega_d, E_k\}$ enters only through
the detuning $\Delta$.
We first analyze how the coherent coupling $H_I$ should be chosen to achieve the most negative value of the coherent contribution $\mathcal{Q}_c$ for fixed $H_0$ and dissipators ${D_i}$.
Replacing $g$ by $p_c$ as an independent control parameter, the optimization naturally separates into two independent steps, whose order can be exchanged.
(i) For a fixed detuning $\Delta$, one maximizes the factor $p_c r^2$.
(ii) For a fixed $p_c$, one maximizes the factor
$\left(\gamma^2 - \Delta^2\right)/\left(\gamma^2 + \Delta^2\right)$.
The first step is equivalent to maximizing the steady-state coherence
$|\langle \Phi_0 | \rho_0 | \Phi_1 \rangle|$, which is achieved when 
$p_c$
is chosen such that $r = r_0/2$.
The second step is naturally realized at $\Delta = 0$, corresponding to the resonant condition.
As a result, we obtain
\begin{equation}
\mathcal{Q}_c^{\min}
=
- \ln \frac{P_0}{P_1}\,
\frac{p_I\, r_0}{4 \gamma},
\end{equation}
where $p_I$ is defined through the limit
$\rho_I = \lim_{g \to \infty} \rho_0$.
In this limit 
\begin{equation} \label{DrhoI}
\mathcal{D}(\rho_I) = \frac{1}{2} p_I r_0 \left( |\Phi_0\rangle\langle\Phi_0| - |\Phi_1\rangle\langle\Phi_1| \right), \end{equation} 
and 
\begin{equation} \label{rhoId}
\langle \Phi_0 | \rho_I | \Phi_0 \rangle = \langle \Phi_1 | \rho_I | \Phi_1 \rangle . 
\end{equation}
Here we also note that the diagonal part of the steady state can be written as
\begin{equation}\label{rhodzw}
\rho_d
=
\frac{p_I}{p_I + p_c}\, \tau
+
\frac{p_c}{p_I + p_c}\, \rho_I .
\end{equation}

 It is clear that the interaction Hamiltonian $H_I$ that minimizes $\mathcal{Q}_c$ does not
necessarily minimize  $\mathcal{Q}_d$, and therefore does not
automatically minimize the total thermodynamic uncertainty $\mathcal{Q}$.
After replacing $g$ by $p_c$, we set $\Delta = 0$, which does not affect the $\mathcal{Q}_d$ part.
Further parameterizing $p_c$ by $r$, the total uncertainty $\mathcal{Q}$ can be expressed as a
quadratic function of $r$.
For the specific examples we studied, this quadratic function has a positive second-order coefficient, 
and its minimum corresponds to an interaction strength satisfying
$4 g^2 \le \gamma p_I$, or equivalently $0 < p_c \le p_I$
—
\textit{we conjecture that this holds generically for the class of models under consideration}.
Specifically, the minimal uncertainty is given by
\begin{equation}\label{EqQmin}
\mathcal{Q}^{\min}
=
\ln \frac{P_0}{P_1}\,
\frac{P}{r_0}
\left[
1
+
\frac{(\mathcal{A}_1 r_0 + \mathcal{B})^2}{4 \mathcal{A}_1}
\right],
\end{equation}
where
 $\mathcal{A}_1 = \mathcal{A} - \frac{p_I}{P \, \gamma}$,
 with $- \frac{p_I}{P \, \gamma}$ coming from $\mathcal{Q}_c$, and $\mathcal{A}$ and $\mathcal{B}$ coming
from $\mathcal{Q}_d$, determined by the model's set of dissipators $\{D_i\}$.
The corresponding extremum occurs at
\begin{equation}\label{rg}
r = \frac{r_0}{2} - \frac{\mathcal{B}}{2 \mathcal{A}_1}.
\end{equation}

To examine the violation of the classical bound, we focus on the reversible
(Carnot) limit, corresponding to $r_0 \to 0$, where the entropy production rate
vanishes.
At $r_0 = 0$, one has $\mathcal{Q}_d = 2$ and $\mathcal{Q}_c = 0$.
In the vicinity of this limit, the first corrections to both contributions appear
at second order in $r_0$.
Consequently, if the quadratic coefficient of $\mathcal{Q}$ in the expansion around
$r_0 = 0$ becomes negative, the total thermodynamic uncertainty falls below the
classical bound.
More specifically, the condition for this behavior at the optimal coupling can be
obtained from Eq.~(\ref{EqQmin}) and reads
\begin{equation}\label{Qleq2}
\lim_{r_0 \to 0}
\frac{(\mathcal{A}_1 r_0 + \mathcal{B})^2}{4 \mathcal{A}_1 r_0^2}+\frac{1}{3 P^2}
<0.
\end{equation}
If the minimal value of the left-hand side, obtained by varying all dissipative parameters ${p_i, R_i}$, satisfies this inequality, then for the given connectivity between the system eigenstates and the reservoirs, the model necessarily allows a regime in which the thermodynamic uncertainty falls below the classical bound.

\section{Physical examples} \label{SecExamp}

The results derived in the previous section show that, for the class of models considered
here, the thermodynamic uncertainty is fully determined by the dissipator
$\mathcal{D}$, the coherent coupling strength $g$, and the detuning $\Delta$.
Its evaluation therefore follows a well-defined and systematic procedure.
Specifically, the dissipator $\mathcal{D}$ determines the decoherence rate $\gamma$,
which, together with $g$ and $\Delta$, fixes the effective coherent rate $p_c$
[see Eq.~(\ref{pc})].
By solving the linear equations associated with $\mathcal{D}$, namely
$\mathcal{D}(\tau) = 0$ and Eqs.~(\ref{DrhoI})–(\ref{rhoId}),
one obtains the steady states $\tau$ and $\rho_I$ in two limiting cases.
According to Eq.~(\ref{rhodzw}), these states determine the diagonal part of the steady state
$\rho_d$ at finite coupling.
The operator $\xi_d$ is then obtained from $\rho_d$ via Eq.~(\ref{xidzw}).
Finally, the diagonal and coherent contributions to the thermodynamic uncertainty,
$\mathcal{Q}_d$ and $\mathcal{Q}_c$, are obtained from Eqs.~(\ref{QdQc}).

In the main text, we illustrate this general framework using two representative models:
a driven single-qubit model and a two-qubit heat-transport model with time-independent coupling,
while additional examples are presented in Appendix~\ref{Examples}.
%
Although our analysis is formulated within a local master equation, we show in Appendix~\ref{global} that, under resonant conditions as defined in Ref.~\cite{bayona2021thermodynamics}, the corresponding global master equation becomes formally equivalent to the local description considered here. This equivalence, while demonstrated explicitly for the models studied in Ref.~\cite{bayona2021thermodynamics}, can be naturally extended to the broader class of models considered in the present work.


\subsection{Driven qubit}

\begin{figure}
 \includegraphics[width=3.13cm]{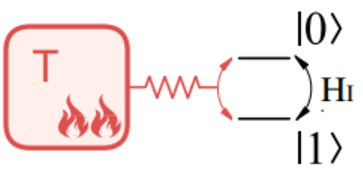}
 \caption{
(Color online)
Schematic illustration of the simplest device converting work into heat.
A  qubit is weakly coupled to a thermal reservoir at temperature $T$ and driven by a time-dependent external field, with the interaction Hamiltonian given in Eq.~(\ref{HIt}).
In this model, the virtual qubit coincides with the physical qubit itself, i.e., 
  $|\Phi_0\rangle$ and $|\Phi_1\rangle$ correspond to the qubit eigenstates $|0\rangle$ and $|1\rangle$, respectively.
 }
 \label{FigModel1_1qubit}
\end{figure}

As a first example, we consider the simplest analytically solvable driven quantum system,
namely a single qubit weakly coupled to a thermal reservoir and subject to a coherent
external field \cite{JPA2021TUR}, as schematically shown in Fig.~\ref{FigModel1_1qubit}.
The bare system Hamiltonian is $H_0=\frac{\omega}{2}\sigma_z$.
In this model, the virtual qubit coincides with the physical qubit itself, such that
$|\Phi_0\rangle=|0\rangle$ and $|\Phi_1\rangle=|1\rangle$.
The qubit is driven by an external field through the interaction Hamiltonian given in Eq.~(\ref{HIt}).
Upon choosing the rotating-frame generator
$G =-   \frac{\omega_d}{2}\sigma_z  $, the dynamics reduces to an equivalent time-independent
coupling problem.
The dissipator associated with the thermal reservoir at temperature $T$ takes the form of Eq.~(\ref{Di}),
with jump operator $\Gamma = |1\rangle\langle 0|$, coupling strength $p$, and thermal occupation factor
$R = [1+\exp(\omega/T)]^{-1}$.
The corresponding decoherence rate is given by $\gamma = p/2$.
The transition from $|0\rangle$ to $|1\rangle$ can occur via the emission of a single quantum ($n=1$)
into the reservoir.

 In the absence of driving ($g=0$), the system relaxes to a thermal equilibrium steady state,
\begin{equation}\label{tau}
\tau = R\,|0\rangle\langle 0| + \bar R\,|1\rangle\langle 1| ,
\end{equation}
with population probabilities $P_0 = R$ and $P_1 = \bar R$,
and hence $r_0 = R - \bar R$ and $P=1$.
Using the properties in Eqs.~(\ref{DrhoI}) and~(\ref{rhoId}), the steady-state limit is found to be
\[
\rho_I = \frac{1}{2} \left( |0\rangle\langle 0| + |1\rangle\langle 1| \right),  
\]
and $p_I = p$.
The diagonal part of the steady state, $\rho_d$, can be
expressed as a convex combination of $\tau$ and $\rho_I$, with weights determined
by the ratio $p_I/p_c$.
The diagonal part of the steady state, $\rho_d$, can be expressed as a convex
combination of $\tau$ and $\rho_I$, with weights determined by the ratio
$p_I/p_c$, where $p_c$ is defined in Eq.~(\ref{pc}).
As a result, the diagonal operator $\xi_d$ appearing in Eq.~(\ref{xidzw})
and its trace can be obtained explicitly, yielding
\[
\mathrm{Tr}\,\xi_d = \frac{1}{p r_0}\left( r + \frac{1}{r} \right).
\]

 \begin{figure}
 \includegraphics[width=4cm]{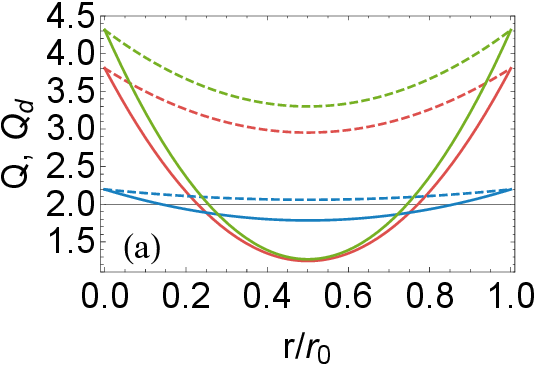}
  \includegraphics[width=4cm]{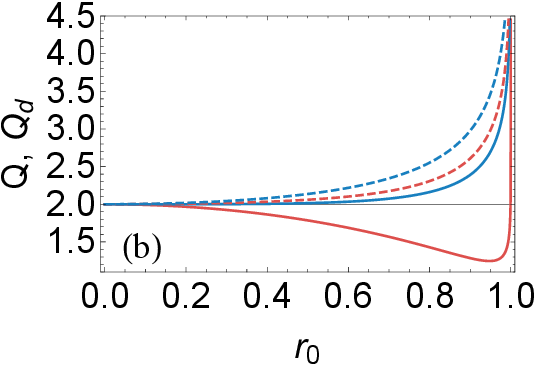}
\caption{
(Color online) 
Thermodynamic uncertainty of the driven qubit model. 
(a) $\mathcal{Q}$ (solid lines) and its diagonal contribution $\mathcal{Q}_d$ (dashed lines) as functions of $r/r_0$ at $\Delta=0$. Red, blue, and green curves correspond to $r_0 = 0.947$, $0.5$, and $0.97$, respectively. 
(b) $\mathcal{Q}$ (solid lines) and $\mathcal{Q}_d$ (dashed lines) as functions of $r_0$ at $\Delta=0$, with $r = 0.5\, r_0$ (red) and $r = \frac{3-\sqrt{5}}{6}\, r_0$ (blue).
}\label{Fig1qubit}
\end{figure}

At this point, we have obtained an analytical expression for the thermodynamic
uncertainty of the model.
Choosing $\Delta = 0$, which corresponds to the condition of minimal uncertainty,
the diagonal and coherent contributions can be expressed in terms of $r$ and $r_0$
as
\begin{eqnarray}
\mathcal{Q}_{d}
&=&
\ln\frac{1+r_0}{1-r_0} \frac{1}{r_0}
\bigl[ 1 - r ( r_0 - r ) \bigr],\nonumber \\
\mathcal{Q}_{c}
&=&
-
\ln\frac{1+r_0}{1-r_0} \frac{2}{r_0}
\, r ( r_0 - r ).
\end{eqnarray}
For a fixed value of $r_0$, both contributions attain their minimum at
$r = r_0/2$.
Furthermore, expanding the total thermodynamic uncertainty $\mathcal{Q}$ around
$r_0 = 0$, the coefficient of the second-order term [the
left-hand side of inequality~(\ref{Qleq2})] is found to be
$-3/4 + 1/3 = -5/12 < 0$.
This result demonstrates that, for the present model, the bound
$\mathcal{Q} < 2$ can indeed be achieved.
 

Figure~\ref{Fig1qubit} illustrates the thermodynamic uncertainty $\mathcal{Q}$ and
its diagonal (classical) contribution $\mathcal{Q}_d$ as functions of $r/r_0$
and $r_0$.
Here, the parameter $r_0$ is determined by the energy splitting of the qubit and
the temperature of the thermal reservoir, while the ratio $r/r_0$ characterizes
the relative strength between the coherent driving amplitude $g$ and the
decoherence rate $p$.
Since the thermodynamic uncertainty is an even function of $r_0$, we present only
the results for $r_0>0$.
In this regime, the time-dependent external field extracts work from an 
negative-temperature reservoir through the driven qubit.
In contrast, the case $r_0<0$ corresponds to a situation where work is performed on
the qubit by the external drive and subsequently dissipated into the thermal
reservoir as heat.

Clearly, both $\mathcal{Q}$ and its diagonal contribution $\mathcal{Q}_d$ are minimized at $r = r_0/2$ for a given $r_0$.
The minimum of $\mathcal{Q}$ occurs at $r = r_0/2$ when $r_0 \simeq 0.947$.
The diagonal part $\mathcal{Q}_d$ always satisfies the classical bound, $\mathcal{Q}_d \ge 2$.
By contrast, the total thermodynamic uncertainty $\mathcal{Q}$ can fall below the classical limit within a finite interval centered at $r = r_0/2$.
This interval broadens as $r_0$ decreases, while the magnitude of the violation simultaneously decreases for $r_0 < 0.947$.
In the limit $r_0 \to 0$, the boundaries of this interval are determined by the condition $3 r (r_0 - r) > r_0^2/3$, yielding $\frac{3-\sqrt{5}}{6} r_0 < r < \frac{3+\sqrt{5}}{6} r_0$.
In the reversible limit, the classical bound on the thermodynamic uncertainty is most susceptible to violation by the interaction Hamiltonian $H_I$, yet the extent of this violation vanishes continuously, with $\mathcal{Q} \to 2^{-}$.

\subsection{Two-qubit heat transport }

 \begin{figure}
 \includegraphics[width=6cm]{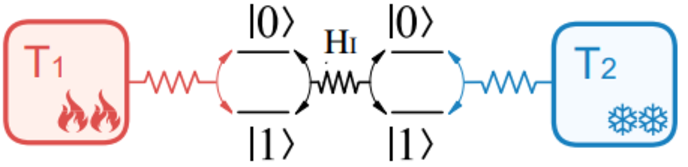}
 \caption{
(Color online)
Schematic illustration of heat transport mediated by two interacting qubits.
Each qubit is weakly coupled to its own thermal reservoir at temperatures $T_1$ and $T_2$, respectively.
The two qubits are coupled through a time-independent interaction of the form given in Eq.~(\ref{HIt}),
corresponding to $\omega_d = 0$.
In this model, the virtual qubit is formed by the two degenerate energy levels of the composite system,
namely $|\Phi_0\rangle = |01\rangle$ and $|\Phi_1\rangle = |10\rangle$.
 }
 \label{FigModel2_2qubit}
\end{figure}

We now consider a system consisting of two resonant qubits, each coupled to its own
thermal reservoir, which together mediate heat transport between two heat baths \cite{EPL2014local,ACMP2015nonequilibrium,
EPL2016perturbative}.
The bare system Hamiltonian reads
\(
H_0 = \frac{\omega}{2}\,\sigma_1^z + \frac{\omega}{2}\,\sigma_2^z ,
\)
where the subscripts label the two qubits.
In this model, the virtual qubit is formed by two degenerate energy levels of the
composite system, namely $|\Phi_0\rangle = |01\rangle$ and $|\Phi_1\rangle = |10\rangle$.
The two qubits are coupled via the interaction given in Eq.~(\ref{HIt}), with
$\omega_d = 0$, such that the detuning is fixed to $\Delta = 0$.

The action of the two thermal reservoirs on the system is described by two
dissipators of the form given in Eq.~(\ref{Di}),
with coupling strengths $p_i$ and thermal occupation factors
$R_i = [1+\exp(\omega/T_i)]^{-1}$.
The corresponding jump operators are $\Gamma_i = \sigma_i^{-}$,
which can equivalently be expressed in the energy eigenbasis of $H_0$.
The decoherence rate of the virtual qubit is $\gamma = (p_1 + p_2)/2$.
The dissipators induce two transition paths between $|01\rangle$ and $|10\rangle$, namely $|01\rangle \to |00\rangle \to |10\rangle$ and $|01\rangle \to |11\rangle \to |10\rangle$, with $n_1 = 1$ and $n_2 = -1$ for both.

In the absence of the interaction $H_I$, each qubit equilibrates with its own reservoir,  
and the steady state of the system is given by
\[
\tau = \tau_1 \otimes \tau_2,
\]
where $\tau_1$ and $\tau_2$ are the thermal equilibrium states of the individual qubits,  
each of the form in Eq.~(\ref{tau}), with excited-state populations $R_1$ and $R_2$, respectively.  
Consequently, the initial population difference of the virtual qubit reads
\(
r_0 = R_1 \bar R_2 - \bar R_1 R_2 = R_1 - R_2.
\)
Using Eqs.~(\ref{DrhoI}) and~(\ref{rhoId}), the state $\rho_I$ is found to be
\[
\rho_I = \frac{1}{2}d \left( |01\rangle\langle 01| + |10\rangle\langle 10| \right) + d_{0} |00\rangle\langle 00| + d_{1} |11\rangle\langle 11|,
\]
with coefficients
\begin{eqnarray}
d &=& \frac{(p_1 R_1 + p_2 R_2)(p_1 \bar R_1 + p_2 \bar R_2)}{2 \gamma^2},
\nonumber \\
d_0 &=& \frac{p_1 R_1 + p_2 R_2}{p_1 \bar R_1 + p_2 \bar R_2} \,  \frac{d}{2},  \quad
d_1 = 1 -  d - d_0. \nonumber 
\end{eqnarray}
and $p_I =  {p_1 p_2}/{\gamma}$.
The diagonal part of the steady state, $\rho_d$, can be written as a convex mixture of $\tau$ and $\rho_I$, where the mixing coefficients are set by the ratio $p_I / p_c$.
One can further assume
\[
\xi_d = k \bigl( |01\rangle\langle 01| + |10\rangle\langle 10| \bigr) 
       + k_0 |00\rangle\langle 00| 
       + k_1 |11\rangle\langle 11|,
\]
and substitute it into Eq.~(\ref{xidzw}). 
The three coefficients $k$, $k_0$, and $k_1$ can be obtained analytically; 
here we omit the explicit expressions for brevity.

With the above results, the thermodynamic uncertainty can be obtained in a closed analytical form,
with the diagonal and coherent contributions reads
\begin{eqnarray}
\mathcal{Q}_d
&=&
\ln\frac{P_0}{P_1}\,
\frac{P}{r_0}
\Bigl[
1 + (r_0 - r)\bigl(\mathcal{A} r + \mathcal{B}\bigr)
\Bigr], \nonumber \\
\mathcal{Q}_c
&=&
\ln\frac{P_0}{P_1}\,
\frac{P}{r_0}\,
\frac{2\,\mathcal{B}}{r_0} \,(r_0 - r)\, r ,
\end{eqnarray}
where the coefficients are given by
\[
\mathcal{A}
=
- \frac{2 (p_1^2 + p_2^2)}{(p_1 + p_2)^2 P},
\qquad
\mathcal{B}
=
- \frac{2 p_1 p_2\, r_0}{(p_1 + p_2)^2 P}.
\]
In this case, the coefficient $\mathcal{A}_1$ appearing in Eq.~(\ref{EqQmin}) is given by
\(
\mathcal{A}_1 = \mathcal{A} + {2\,\mathcal{B}}/{r_0}=-2/P.
\)

 \begin{figure}
 \includegraphics[width=4cm]{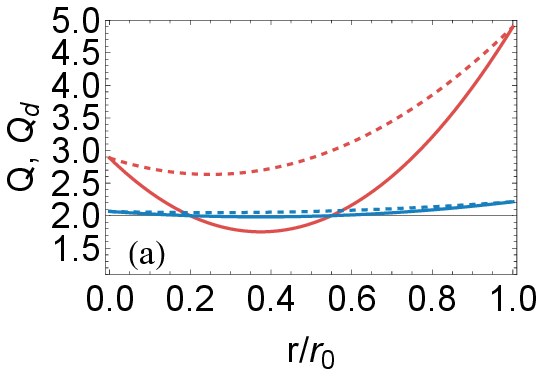}
  \includegraphics[width=4cm]{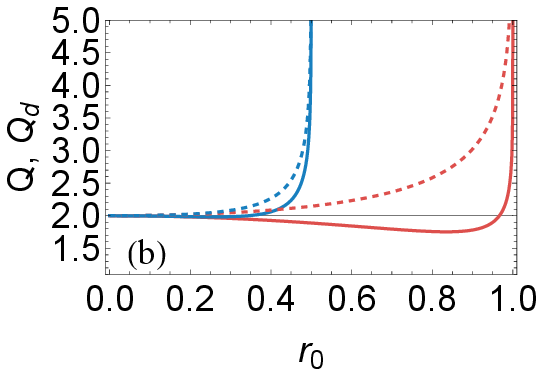}
\caption{(Color online)
Thermodynamic uncertainty in the two-qubit heat-transport model with $p_1=p_2$.
(a) The total thermodynamic uncertainty $\mathcal{Q}$ (solid lines) and its diagonal contribution
$\mathcal{Q}_d$ (dashed lines) as functions of $r/r_0$.
The red curve corresponds to 
$R_{1,2}=(1\pm r_0)/2$ with $r_0=0.835$,
while the blue curve corresponds to
$R_1=1/2$ and $R_2=1/2-r_0$ with $r_0=0.259$.
(b) $\mathcal{Q}$ (solid lines) and $\mathcal{Q}_d$ (dashed lines) as functions of $r_0$,
where $r=3r_0/8$.
The red and blue curves correspond to
$R_{1,2}=(1\pm r_0)/2$ and
$R_1=1/2$, $R_2=1/2-r_0$, respectively.}
\label{Fig2qubit}
\end{figure}

For fixed reservoir occupations $\{R_i\}$, which determine the parameters $P$ and
$r_0$ appearing above, the minimum of the thermodynamic uncertainty can be obtained
through two independent optimization steps.
First, the system--reservoir couplings must satisfy the symmetric condition
$p_1 = p_2$.
Second, the internal coupling strength $g$ is chosen such that the parameter $r$
satisfies Eq.~(\ref{rg}); in particular, for $p_1 = p_2$ the optimal value is
$r = 3 r_0 / 8$.
Under these conditions, the minimal thermodynamic uncertainty takes the simple form
\begin{equation}\label{Q2qubit}
\mathcal{Q}
=
\ln\frac{P_0}{P_1}\,
\frac{P}{r_0}
\left(
1 - \frac{25\, r_0^2}{32\, P}
\right).
\end{equation}

  For a fixed value of $r_0$, one can further minimize Eq.~(\ref{Q2qubit}) with respect
to the parameter $P$, which leads to two distinct cases depending on the allowed
range of reservoir occupations.
Owing to the symmetry of the problem, we restrict ourselves to $r_0>0$.
In the first case, the occupations satisfy $R_i \in (0,1/2)$, which restricts
$r_0$ to the interval $[0,1/2)$ and corresponds to reservoirs at finite temperatures.
In this case, the minimum of $\mathcal{Q}_{\mathrm{2qubit}}$ is attained at
$R_1 \to 1/2$ and $R_2 = 1/2 - r_0$.
In the second case, the occupations satisfy $R_i \in (0,1)$, thereby allowing for
reservoirs with negative effective temperatures.
Here, the minimum of $\mathcal{Q}_{\mathrm{2qubit}}$ is achieved for the symmetric
choice $R_{1,2} = 1/2 \pm r_0/2$.

Figure~\ref{Fig2qubit} illustrates the behavior of the total thermodynamic uncertainty
$\mathcal{Q}$ and its diagonal contribution $\mathcal{Q}_d$ as functions of $r/r_0$
and $r_0$ for these two optimal parameter choices.
When negative-temperature reservoirs are allowed, the global optimum occurs at
$r_0 \simeq 0.835$, yielding $\mathcal{Q} \simeq 1.76$, whereas restricting the
reservoirs to finite temperatures leads to an optimal value $r_0 \simeq 0.259$
with $\mathcal{Q} \simeq 1.98$.
At the respective optimal values of $r_0$, in the finite-temperature case both the
total uncertainty $\mathcal{Q}$ and its diagonal contribution $\mathcal{Q}_d$ remain
close to the classical bound $\mathcal{Q}=2$, while in the negative-temperature case
$\mathcal{Q}$ exhibits a much stronger dependence on the interaction strength, as
quantified by $r/r_0$.
Despite this quantitative difference, the range of $r/r_0$ for which $\mathcal{Q}<2$
is similar in both cases, approximately between $0.2$ and $0.55$.
Finally, in the limit $r_0 \to 0$ the two cases coincide and share the same leading
behavior,
\(
\mathcal{Q} \simeq 2 - \frac{11\, r_0^2}{24}.
\)

\section{Conclusion} \label{SecConcl}

We have presented a unified analysis of the TUR for a broad class of quantum thermal machines whose operation is enabled by coherent coupling within a virtual qubit, defined by two eigenstates of the bare system Hamiltonian.
In this class of models, steady-state coherence is confined to the virtual-qubit subspace, which allows for a transparent identification of classical contributions associated with the diagonal elements of the density matrix and genuinely quantum contributions arising from off-diagonal coherences.

We developed a systematic procedure to evaluate the thermodynamic uncertainty. The central step consists in solving three coupled linear equations determined by the dissipators. Within this framework, we identified the optimal coupling conditions and established a clear criterion for surpassing the classical thermodynamic uncertainty bound in the vicinity of the reversible limit.
The steady-state currents and the associated entropy production can always be faithfully reproduced by an effective classical Markovian process. In contrast, the quantum nature of the dynamics manifests itself in second-order current fluctuations, which acquire an additional purely quantum contribution originating from steady-state coherence. This observation naturally leads to a decomposition of the thermodynamic uncertainty into a diagonal (classical) part, obeying the standard thermodynamic uncertainty bound, and a coherent part, which can become strictly negative under resonant conditions.
The coupling strength minimizing the coherent contribution to the thermodynamic uncertainty simultaneously maximizes the steady-state coherence within the virtual-qubit subspace, thereby revealing a direct connection between coherence as a quantum resource and the precision of energy transport.

Our framework applies equally to autonomous thermal machines with time-independent couplings and to driven quantum thermal machines. Although our analysis is based on a local master equation, we showed that, under resonant conditions as defined in Ref.~\cite{bayona2021thermodynamics}, the master equation obtained within a global approach becomes formally equivalent to our description.

It would be interesting to extend the present analysis to more general scenarios. For instance, systems exhibiting coherence among multiple energy levels go beyond the virtual-qubit picture considered here. While previous works have shown that such coherences may also lead to violations of the thermodynamic uncertainty bound \cite{JPA2021TUR}, a unified treatment of these models may provide deeper and more general insights. Moreover, different types of quantum correlations induced by coherence—such as entanglement, nonlocality, and quantum steering—play hierarchical roles in quantum information tasks 
\cite{RevModPhys.81.865,RMP2012Vedral,RMP2014bell,RMP2020steer}. Clarifying whether and how these different quantum correlations contribute to the thermodynamic uncertainty in a hierarchical manner constitutes an interesting direction for future research at the interface of quantum information and quantum thermodynamics.
As a concrete example, temporal quantum correlations characterized by the Leggett–Garg inequality \cite{leggett1985quantum} share strong conceptual similarities with fluctuation measures, and their relation to power fluctuations has been discussed recently \cite{Friedenberger2017,Blattmann2017,miller2018leggett}.

\bibliography{QuantumTUR}

\begin{acknowledgments}
This work was supported by the National Natural Science
 Foundation of China (No. 11675119)
\end{acknowledgments}

\onecolumngrid  

\appendix

\section{Jump Operators Acting on the Virtual Qubit}\label{Gamma01}

In this Appendix, we analyze the structure of the jump operators acting on the
virtual-qubit subspace spanned by $\{|\Phi_0\rangle, |\Phi_1\rangle\}$.
We require that the dissipative dynamics does not generate coherences outside
this subspace.
The most general jump operator that can couple to the virtual-qubit subspace spanned by $\{|\Phi_0\rangle, |\Phi_1\rangle\}$ can be decomposed as \begin{eqnarray} \Gamma &=& a_k |\Phi_0\rangle\langle \Phi_k| + b_l |\Phi_1\rangle\langle \Phi_l| \nonumber \\ &\ & + c_m |\Phi_m\rangle\langle \Phi_0| + d_n |\Phi_n\rangle\langle \Phi_1| + \Gamma_o , \end{eqnarray} where the indices satisfy $k \neq 0$, $l \neq 1$, and $m,n \neq 0,1$.
The contribution $\Gamma_o$ acts entirely outside the virtual-qubit subspace and
contains none of the bra or ket states appearing in the first four terms.
%
%
%
Owing to the symmetry of the dissipator, it suffices to examine their action
on the operator $|\Phi_0\rangle\langle\Phi_1|$.
First, the anticommutators
$\{\Gamma \Gamma^{\dagger},\,|\Phi_0\rangle\langle\Phi_1|\}$ and
$\{\Gamma^{\dagger}\Gamma,\,|\Phi_0\rangle\langle\Phi_1|\}$
consist of terms of the form
$|\Phi_p\rangle\langle\Phi_1|$ and $|\Phi_0\rangle\langle\Phi_q|$.
We then analyze the action of the jump operators by distinguishing two cases.

\medskip
\noindent
\textbf{(i)} $\Gamma$ contains either $|\Phi_0\rangle\langle\Phi_1|$
or $|\Phi_1\rangle\langle\Phi_0|$.
Without loss of generality, we assume that the former is present,
i.e., we take $k=1$ and $a_1 \neq 0$.
In this situation, one must have $d_n = 0$, $l \neq 0,1$, and $m \neq 0,1$.
The action of such a jump operator on $|\Phi_0\rangle\langle\Phi_1|$ reads
\begin{eqnarray}
\Gamma^{\dagger} |\Phi_0\rangle\langle\Phi_1| \Gamma
&=& a_1^{*} b_l \, |\Phi_1\rangle\langle\Phi_l|, \\
\Gamma |\Phi_0\rangle\langle\Phi_1| \Gamma^{\dagger}
&=& c_m a_1^{*}\, |\Phi_m\rangle\langle\Phi_0|.
\end{eqnarray}
These terms cannot cancel the contributions from the anticommutator part,
and therefore one must require $b_l = c_m = 0$.
Similarly, when $l=0$ and $b_0 \neq 0$, one obtains
$a_k = c_m = d_n = 0$.

\medskip
\noindent
\textbf{(ii)} $\Gamma$ contains neither $|\Phi_0\rangle\langle\Phi_1|$
nor $|\Phi_1\rangle\langle\Phi_0|$.
In this case, $k,l,m,n \neq 0,1$.
A similar calculation yields
\begin{eqnarray}
\Gamma^{\dagger} |\Phi_0\rangle\langle\Phi_1| \Gamma
&=& a_k^{*} b_l \, |\Phi_k\rangle\langle\Phi_l|, \\
\Gamma |\Phi_0\rangle\langle\Phi_1| \Gamma^{\dagger}
&=& c_m d_n^{*} \, |\Phi_m\rangle\langle\Phi_n|.
\end{eqnarray}
From this, one concludes that $a_k b_l = 0$ and $c_m d_n = 0$.
In both cases, the corresponding anticommutators
\[
\{ \Gamma\Gamma^{\dagger},\,|\Phi_0\rangle\langle\Phi_1| \}
\quad \text{and} \quad
\{ \Gamma^{\dagger}\Gamma,\,|\Phi_0\rangle\langle\Phi_1| \}
\]
are proportional to $|\Phi_0\rangle\langle\Phi_1|$.

\medskip
Collecting the above results, we find that the jump operators acting on the
virtual-qubit states can only take the following forms, or their Hermitian
conjugates:
\begin{eqnarray}
\Gamma_i &=& a_1 |\Phi_0\rangle\langle \Phi_1| + \Gamma_o , \nonumber\\
\Gamma_i &=& a_k |\Phi_0\rangle\langle \Phi_k|
            + c_m |\Phi_m\rangle\langle \Phi_0| + \Gamma_o, \\
\Gamma_i &=& a_k |\Phi_0\rangle\langle \Phi_k|
            + d_n |\Phi_n\rangle\langle \Phi_1| + \Gamma_o, \nonumber\\
\Gamma_i &=& b_l |\Phi_1\rangle\langle \Phi_l|
            + d_n |\Phi_n\rangle\langle \Phi_1| + \Gamma_o, \nonumber      
\end{eqnarray}
Here $k,l,m,n \neq 0,1$, and $\Gamma_o$ denotes operators that induce transitions
entirely outside the virtual-qubit subspace.

\section{System-Reservoir Model} 

The total Hamiltonian of the system and the modes in each reservoir interacting with the system can be written as
\begin{eqnarray}\label{HBS}
H = H_s + \sum_i \omega_i \, b_i^\dagger b_i + V ,
\end{eqnarray}
where $b_i$ annihilates a photon of frequency $\omega_i$ in the $i$th reservoir, and $V$ denotes the system--reservoir coupling.
Only the rotating-wave components of the coupling contribute to the dissipative terms in the master equation~(\ref{MasterEq}), namely,
\begin{eqnarray}\label{VBS}
V = \sum_{k,l,i} \delta_{E_l - E_k, \omega_i} \, g_{kl}^{(i)}
\left(
|\Phi_l\rangle \langle \Phi_k| b_i
+
|\Phi_k\rangle \langle \Phi_l| b_i^\dagger
\right).\ \ \ 
\end{eqnarray}
In practice, each reservoir contains a large number of additional harmonic modes that are not explicitly included in the Hamiltonian~(\ref{HBS}) and are assumed to remain in thermal equilibrium.
Weak interactions among these modes induce dissipative relaxation, ensuring that any mode driven out of equilibrium relaxes back to thermal equilibrium.

The operator that counts the net number of photons transferred from the system to the $i$-th reservoir during the time interval $[0,t]$ is defined as
\begin{eqnarray}
    \tilde{N}_i(t) = N_i(t) - N_i ,
\end{eqnarray}
with $N_i = b_i^\dagger b_i$ and
$A(t) = \mathcal{U}^{\dagger}(t) A \mathcal{U}(t)$ denoting operators in the Heisenberg representation.
Here $\mathcal{U}(t)=\mathcal{T}\exp\!\left(-i \int_{0}^{t} H\, \mathrm{d}t\right)$ is the time-evolution operator, and $\mathcal{T}$ denotes time ordering.
The corresponding photon number current and its variance are defined in the long-time limit as
\begin{eqnarray}
    J_i &=& \lim_{t \to +\infty} \frac{1}{t}\, \mathrm{Tr}\big[ \tilde{N}_i(t)\, \rho(0)\big],\\
    \mathrm{Var}(J_i) &=& \lim_{t \to +\infty} \frac{1}{t} 
    \left\{ \mathrm{Tr}\big[ \tilde{N}_i^2(t)\, \rho(0)\big] - (J_i t)^2 \right\},
\end{eqnarray}
where $\rho(0)$ denotes the total density matrix of the system and reservoirs at the initial time $t=0$.
These quantities can equivalently be obtained from the cumulant generating function by taking derivatives with respect to $i \chi$,
\begin{eqnarray}\label{CGF}
    C_i(\chi) =  \lim_{t \to +\infty} \frac{1}{t}\, 
    \ln \left\{ \mathrm{Tr}
    \big[   e^{  i \frac{\chi}{2} N_i}\, \mathcal{U}(t)\, e^{- i \frac{\chi}{2} N_i}   \rho  (0)  e^{- i \frac{\chi}{2} N_i}\, \mathcal{U}^{\dagger}(t)\, e^{  i \frac{\chi}{2} N_i}    \big] \right\}
    =  \lim_{t \to +\infty} \frac{1}{t}\, 
    \ln \left\{ \mathrm{Tr}\big[ \rho_i^{\chi}(t)\big] \right\}.
\end{eqnarray}
Here the counting-field–dependent density matrix is defined as
\begin{eqnarray}
    \rho_i^{\chi}(t) = \mathcal{U}_i^{-\chi}(t)\, \rho(0)\, \mathcal{U}_i^{\chi \dagger}(t),
\end{eqnarray}
with the modified evolution operators given by
\begin{eqnarray}
    \mathcal{U}_i^{-\chi}(t) &=& e^{  i \frac{\chi}{2} N_i}\, \mathcal{U}(t)\, e^{- i \frac{\chi}{2} N_i},\\
    \mathcal{U}_i^{\chi \dagger}(t) &=& e^{- i \frac{\chi}{2} N_i}\, \mathcal{U}(t)\, e^{  i \frac{\chi}{2} N_i}.
\end{eqnarray}
The $n$-th cumulant of the photon number current is then obtained by taking the $n$-th derivative of $C_i(\chi)$ with respect to $i\chi$ at $\chi=0$.

\subsection{Removing Time Dependence}\label{NoTime}

Now, we show that a model with explicit time-dependent driving can always be mapped onto an equivalent model with stationary couplings. 
To this end, we introduce a ``rotating'' generator
\begin{eqnarray}
I = \sum_{i} \Omega_i\, b_i^{\dagger} b_i + G,
\end{eqnarray}
where $G = \sum_{k} e_k\, |\Phi_k\rangle \langle \Phi_k|$.
Here we require that, for any pair of states satisfying $E_l - E_k = \omega_i$, the coefficients obey
$e_l - e_k = \Omega_i$, and furthermore $e_0 - e_1 = \omega_d$.
Such a set of parameters $\{ \Omega_i, e_k \}$ always exists and is not unique.
In particular, for the resonant case $E_0 - E_1 = \omega_d$, a convenient and simplest choice is given by $G = H_0$ and $\Omega_i = \omega_i$.

The density matrix in the rotating frame is given by
\begin{eqnarray}
\rho^{\rm rot}(t) = \mathcal{R}(t)\, \rho(t)\, \mathcal{R}^{\dagger}(t),
\qquad
\mathcal{R}(t)= \exp(- i I t).
\end{eqnarray}
Its time evolution is governed by
\begin{eqnarray}\label{Seqrot}
\dot{\rho}^{\rm rot}(t) = - i \big[ H^{\rm rot}, \rho^{\rm rot}(t) \big],
\end{eqnarray}
with the effective Hamiltonian in the rotating frame given by
\begin{eqnarray}
H^{\rm rot} = H\big|_{t=0} + I ,
\end{eqnarray}
which is time independent.
Correspondingly, the full time-evolution operator can be factorized as
\begin{eqnarray}
\mathcal{U}(t)=\mathcal{R}(t)\,\mathcal{U}^{\rm rot}(t),
\qquad
\mathcal{U}^{\rm rot}(t)=\exp(- i H^{\rm rot} t).
\end{eqnarray}
Substituting this decomposition into Eq.~(\ref{CGF}) and using the fact that
$[I, N_i]=0$, one obtains
\begin{eqnarray}\label{CGFrot}
    C_i(\chi) =  \lim_{t \to +\infty} \frac{1}{t}\, 
    \ln \left\{ \mathrm{Tr}
    \big[   e^{  i \frac{\chi}{2} N_i}\, \mathcal{U}^{\rm rot} (t)\, e^{- i \frac{\chi}{2} N_i}   \rho  (0)  e^{- i \frac{\chi}{2} N_i}\, \mathcal{U}^{\rm rot \dagger}(t)\, e^{  i \frac{\chi}{2} N_i}    \big] \right\} .
\end{eqnarray}
This demonstrates that the photon currents and all their cumulants generated by the time-dependent Hamiltonian $H$ are identical to those generated by the time-independent Hamiltonian $H^{\rm rot}$.
Moreover, under the above rotating transformation, the thermal equilibrium states of the reservoirs and the interaction term $V$ remain unchanged, and the resonance conditions between the system energy levels and the reservoir modes are preserved.
As a consequence, within the local approach, the dissipative operators derived from Eq.~(\ref{Seqrot}) are identical to those of the original model.
The reduced density matrix of the system obeys
\begin{eqnarray} \label{MasterRot}
\dot{\rho}_s^{\rm rot} = - i \big[ H_s^{\rm rot}, \rho_s^{\rm rot} \big]+\sum_i D_i(\rho_s^{\rm rot}),
\end{eqnarray}
with the effective Hamiltonian in the rotating frame given by
\begin{eqnarray}
H_s^{\rm rot} = H_s\big|_{t=0} + G ,
\end{eqnarray}
which is time independent.
The same conclusion can alternatively be reached by performing the rotating transformation directly at the level of the master equation~(\ref{MasterEq}).
From Eqs.~(\ref{CGFrot}) and~(\ref{MasterRot}), one can further see that the non-uniqueness in the choice of $I$ and $G$ does not affect the steady state or the heat currents. In particular, among the diagonal terms of $H_s^{\rm rot}$, the only contribution relevant to the results arises from the detuning
$\Delta = E_0 - E_1 - \omega_d$.

\subsection{Relations Among Currents from Different Reservoirs}\label{G0G12}

To investigate the relations among the heat currents associated with different reservoirs, 
we consider three reservoirs, labeled by $0,1,2$.
Along the transition path from $|\Phi_0\rangle$ to $|\Phi_1\rangle$, the associated photon numbers satisfy
$n_0 = 0$ and $n_1 \neq 0$, $n_2 \neq 0$.
To characterize the properties of the corresponding heat currents, we introduce two operators
\begin{eqnarray}
I_0  &=& N_0 + G_0,\\
I_{12} &=& \frac{N_1}{n_1} - \frac{N_2}{n_2} + G_{12},
\end{eqnarray}
  where $G_0=\sum_{k} \epsilon_{k}|\Phi_k\rangle \langle \Phi_k| $ and $G_{12}=\sum_{k}  \varepsilon_{k}|\Phi_k\rangle \langle \Phi_k| $.
  By an appropriate choice of the coefficients $\epsilon_k$ and $\varepsilon_k$, the operators $I_0$ and $I_{12}$ can be constructed as conserved quantities, in the sense that
\begin{eqnarray}
[I_0,\,H]=0, \qquad [I_{12},\,H]=0 .
\end{eqnarray}

The coefficients $\epsilon_k$ are assigned as follows: for any pair of levels coupled to reservoir $0$, i.e., those satisfying
$E_l - E_k = \omega_0$ in the interaction $V$ [cf. Eq.~(\ref{VBS})], we set
$\epsilon_l - \epsilon_k = 1$, whereas for level pairs not coupled to reservoir $0$ we take
$\epsilon_{l'} - \epsilon_{k'} = 0$.
A subset of $\epsilon_k$ is further fixed to zero in order to determine all $\epsilon_k$ uniquely, while remaining consistent with the above constraints.

The coefficients $\varepsilon_k$ in $G_{12}$ can be assigned in a similar, self-consistent manner.
For pairs of levels coupled to reservoir $1$, i.e., those satisfying
$E_l - E_k = \omega_1$ in the interaction $V$ [cf. Eq.~(\ref{VBS})], we set
$\varepsilon_l - \varepsilon_k =  {1}/{n_1}$;
for pairs of levels coupled to reservoir $2$, satisfying
$E_l - E_k = \omega_2$, we set
$\varepsilon_l - \varepsilon_k = - {1}/{n_2}$;
and for level pairs not coupled to either of these two reservoirs we take
$\varepsilon_{l'} - \varepsilon_{k'} = 0$.
Again, a subset of $\varepsilon_k$ is fixed to zero so as to determine all coefficients uniquely.
With this construction, one obtains $\varepsilon_0 = \varepsilon_1$.

Intuitively, in the long-time limit the system relaxes to a steady state, and the variations of the operators
\begin{eqnarray}
\tilde{G}_0(t)   =  G_0(t) -  {G}_0, \qquad
\tilde{G}_{12}(t)   =  G_{12}(t) -  {G}_{12},
\end{eqnarray}
become equivalent to the zero operator.
As a consequence, all moments of $\tilde{N}_0(t)$ vanish, while all moments of
$\frac{1}{n_1}\tilde{N}_1(t)$ and $\frac{1}{n_2}\tilde{N}_2(t)$ are identical.
More rigorously, this result can be demonstrated starting from the equation satisfied by the
counting-field–dependent density matrix, which reads
\begin{eqnarray}\label{Chitotal}
    \dot{\rho}_i^{\chi}(t) = -i   \bigg(
    H_i^{-\chi} \, \rho_i^{\chi}(t)
    - \rho_i^{\chi}(t)\, H_i^{ \chi} 
    \bigg),
\end{eqnarray}
where $H_i^{-\chi}  =e^{  i \frac{\chi}{2} N_i}\, H\, e^{- i \frac{\chi}{2} N_i}$.
The cumulant generating function is determined by the eigenvalue of the
corresponding superoperator,
\begin{eqnarray}
 - i \Big(
 H_i^{-\chi}\, \varrho_i(\chi)
 - \varrho_i(\chi)\, H_i^{\chi}
 \Big)
 = \lambda_i(\chi)\, \varrho_i(\chi),
\end{eqnarray}
where $\lambda_i(\chi)$ denotes the eigenvalue with the largest real part, which governs the long-time behavior of the cumulants.
Making use of the conserved quantities $I_0$ and $I_{12}$, one finds that the
counting-field–dependent total Hamiltonians satisfy
\begin{eqnarray}
H_0^{-\chi} &=&  e^{ - i \frac{\chi}{2} G_0}\, H\, e^{ i \frac{\chi}{2} G_0},\\
H_{1}^{-\chi} &=&  e^{ - i \frac{\chi}{2} G_{12}}\,
H_{2}^{-\frac{n_1}{n_2}\chi}\,
e^{ i \frac{\chi}{2} G_{12}}.
\end{eqnarray}
As a consequence, the corresponding eigenvalues and eigenoperators obey
\begin{eqnarray}
\lambda_0(\chi)= \lambda_0(0), &\qquad&
\varrho_0(\chi) = e^{ - i \frac{\chi}{2} G_0}\, \varrho_0(0)\, e^{- i \frac{\chi}{2} G_0},\\
\lambda_1(\chi) = \lambda_2\!\left(\frac{n_1}{n_2}\chi \right), &\qquad&
\varrho_1(\chi)
= e^{ - i \frac{\chi}{2} G_{12}}\,
\varrho_2\!\left(\frac{n_1}{n_2}\chi \right)\,
e^{ -i \frac{\chi}{2} G_{12}} .
\end{eqnarray}
This completes the proof.

\section{Classical Counterpart}\label{App_Classical}

The action of the right-hand side of the master equation~(\ref{MasterEq})
(i.e., the Liouvillian)
is closed on the operator set
$\{ |\Phi_k\rangle\langle\Phi_k|,\,
   |\Phi_0\rangle\langle\Phi_1|,\,
   |\Phi_1\rangle\langle\Phi_0| \}$.
As a consequence, for stationary coupling the steady state of the master equation
can be written in the form
\begin{equation}
\rho_s = \rho_d + \rho_c ,
\end{equation}
where the diagonal (classical) part is
\begin{equation}
\rho_d = \sum_k q_k\, |\Phi_k\rangle\langle\Phi_k| ,
\end{equation}
and the coherence is confined to the virtual-qubit subspace,
\begin{equation}\label{rhoc}
\rho_c = z\, |\Phi_0\rangle\langle\Phi_1|
       + z^*\, |\Phi_1\rangle\langle\Phi_0| .
\end{equation}
Here the populations satisfy $\sum_k q_k = 1$ and $q_k \geq 0$, and the complex
coherence is parameterized as $z = x - i y$ with real $x$ and $y$.
Substituting this ansatz into the master equation and imposing the steady-state
condition $\dot{\rho}_s = 0$, one obtains
\begin{eqnarray}\label{xy}
    x = \frac{g\, \Delta\, r}{\Delta^2 + \gamma^2}, \qquad
y =- \frac{g\, \gamma\, r}{\Delta^2 + \gamma^2},
\end{eqnarray}
where $r = q_0 - q_1$ denotes the population imbalance of the virtual qubit.
 
With these expressions, the steady-state equation can be rearranged as
\begin{equation}\label{Dcrhod}
- i [H_0,\rho_d] + \mathcal{D}(\rho_d)
= i [H_I,\rho_s] + i [H_0,\rho_c] - \mathcal{D}(\rho_c)
= \frac{p_c}{2}\, r \left(
|\Phi_0\rangle\langle\Phi_0|
-
|\Phi_1\rangle\langle\Phi_1|
\right),
\end{equation}
where the effective rate
\begin{equation}
p_c = \frac{4 g^2 \gamma}{\gamma^2 + \Delta^2}
\end{equation}
naturally emerges.
We choose the dissipator $D_c$ in the Lindblad form
\begin{eqnarray}
D_c(\rho)
= p_c R_c \, \Big(
 \Gamma_c^{\dagger} \rho \Gamma_c
- \frac{1}{2} \{ \Gamma_c \Gamma_c^{\dagger}, \rho \}
\Big)  
+ p_c \bar{R}_c \, \Big(
 \Gamma_c \rho \Gamma_c^{\dagger}
- \frac{1}{2} \{ \Gamma_c^{\dagger} \Gamma_c, \rho \}
\Big),
\end{eqnarray}
with the jump operator
\begin{equation}
\Gamma_c = |\Phi_1\rangle \langle \Phi_0|,
\end{equation}
and symmetric rates
\begin{equation}
R_c = \bar{R}_c = \frac{1}{2}.
\end{equation}
When acting on the diagonal density matrix $\rho_d$, the contribution generated by
$D_c(\rho_d)$ exactly cancels the term given in Eq.~(\ref{Dcrhod}).
As a consequence, the steady state of the master equation
(\ref{MasterEqClassical}) is precisely given by $\rho_d$.

 \section{Full Counting Statistics}\label{FCS}

In this section, we employ the full counting statistics (FCS) formalism to analyze
the master equation~(\ref{MasterEq}) under stationary coupling.
This treatment is equivalent to starting from the microscopic
system--reservoir model described by Eq.~(\ref{Chitotal}) and reducing the influence
of the reservoirs on the system to Lindblad-type dissipators.
For convenience, we redefine the counting-field parameter according to
$i\chi \rightarrow \chi$.
Throughout this section, $\rho$ denotes the reduced density matrix of the system,
while $\rho^{\chi}$ represents the system state modified by the counting field.

\subsection{Perturbative Treatment}

When we focus on the heat current associated with the $i$th reservoir
($n_i \neq 0$), the modified master equation takes the form
\begin{equation}\label{MasterEqChi}
\dot{\rho}^{\chi}
= -i [H_s, \rho^{\chi}] + \mathcal{D}^{\chi}(\rho^{\chi}),
\end{equation}
where the dissipator $\mathcal{D}^{\chi}$ has the same structure as
$\mathcal{D}$ in Eq.~(\ref{MasterEq}), except that the $i$th dissipators $D_i$ is replaced by 
$D_i^{\chi}$,
\begin{eqnarray}\label{Dix}
D_i^{\chi}(\rho)
= p_i R_i
\left(
e^{-\chi}\Gamma_i^{\dagger} \rho \Gamma_i
-\frac{1}{2}\{\Gamma_i \Gamma_i^{\dagger}, \rho\}
\right) 
+ p_i \bar{R}_i
\left(
e^{\chi}\Gamma_i \rho \Gamma_i^{\dagger}
-\frac{1}{2}\{\Gamma_i^{\dagger} \Gamma_i, \rho\}
\right).
\end{eqnarray}
In this formulation, the cumulant generating function introduced in
Eq.~(\ref{CGF}) is given by
\begin{equation}
C_i(\chi)
= \lim_{t \to +\infty} \frac{1}{t}
\ln \mathrm{Tr}  \rho^{\chi} .
\end{equation}
The steady-state current and its variance are expressed as
\begin{equation}
J_i
= \lim_{t \to +\infty} \frac{1}{t} \left.
\partial_\chi
\ln \mathrm{Tr}   \rho^{\chi}
\right|_{\chi=0}, \qquad
\mathrm{Var}(J_i)
= \lim_{t \to +\infty} \frac{1}{t}\left.
\partial_\chi^2
\ln \mathrm{Tr}   \rho^{\chi}  
\right|_{\chi=0}.
\end{equation}
We expand the long-time limit of $\rho^{\chi}$ in powers of $\chi$ as
\begin{equation}
\lim_{t \to +\infty} \rho^{\chi}
= \rho_0 + \chi\, \rho_1 + \chi^2 \rho_2 + \cdots .
\end{equation}
At this stage, the current and its variance are given by
\begin{equation}
J_i = \frac{1}{t} \mathrm{Tr}\,\rho_1, \qquad
\mathrm{Var}(J_i)
= \frac{1}{t}
\left[
2\,\mathrm{Tr}\,\rho_2 - (J_i t)^2
\right].
\end{equation}
On the other hand, in the long-time limit one has
\begin{equation}
\lim_{t \to +\infty} \rho^{\chi}
= e^{\lambda(\chi) t}\, \varrho^{\chi},
\end{equation}
where $\lambda(\chi)$ is the eigenvalue of the Liouvillian
$\mathcal{L}^{\chi}   =-i [H_s, \cdot ] + \mathcal{D}^{\chi}(\cdot  ) $ 
with the largest real part, and
$\varrho^{\chi}$ is the corresponding eigenoperator, satisfying
\begin{equation}
\mathcal{L}^{\chi}\, \varrho^{\chi}
= \lambda(\chi)\, \varrho^{\chi}.
\end{equation}
The eigenoperator $\varrho^{\chi}$ can be expanded in powers of $\chi$ as
 $
\varrho^{\chi} = \sum_{n=0}^{+\infty} \varrho_n\, \chi^n .
$
Here $\varrho_0=\rho_s$ is the steady state of the master equation~(\ref{MasterEq}),
corresponding to the eigenoperator at $\chi=0$, for which $\lambda(0)=0$.
Without loss of generality, we choose the normalization
$\mathrm{Tr}\,\varrho^{\chi}=1$, which implies $\mathrm{Tr}\,\varrho_n=0$
for $n>0$.
Expanding $e^{\lambda(\chi) t}\, \varrho^{\chi}$ in powers of $\chi$ up to second order, one finds
\begin{equation}\label{3order}
\rho_0=\varrho_0, 
\quad 
\rho_1=
t\,\lambda'\,\varrho_0 + \varrho_1, 
\quad 
\rho_2=
\frac{1}{2}
\Big(
t\,\lambda'' + t^2\,\lambda'^2
\Big)\varrho_0
+ t\,\lambda'\,\varrho_1
+ \varrho_2,
\end{equation}
where $\lambda'=\left.\partial_\chi \lambda(\chi)\right|_{\chi=0}$ and
$\lambda''=\left.\partial_\chi^2 \lambda(\chi)\right|_{\chi=0}$.

To determine the steady-state current and its variance explicitly,
we expand the Liouvillian in powers of the counting field as
\begin{equation}
\mathcal{L}^{\chi}
= \sum_{n=0}^{\infty} \frac{1}{n!}\, \chi^{n}\, \mathcal{L}_{n},
\end{equation}
where the individual components are given by
\begin{eqnarray}\label{Ln}
\mathcal{L}_0 \rho
&=& -i [H_s, \rho] + \mathcal{D}(\rho), \\
\mathcal{L}_{n} \rho
&=& - p_i R_i \, \Gamma_i^{\dagger} \rho \Gamma_i
+ p_i \bar{R}_i \, \Gamma_i \rho \Gamma_i^{\dagger},
\qquad n = 1,3,5,\ldots, \\
\mathcal{L}_{n} \rho
&=& p_i R_i \, \Gamma_i^{\dagger} \rho \Gamma_i
+ p_i \bar{R}_i \, \Gamma_i \rho \Gamma_i^{\dagger},
\qquad n = 2,4,6,\ldots .
\end{eqnarray}
The modified master equation then yields the following hierarchy:
\begin{equation}
\dot{\rho}_n
= \sum_{m=0}^{n} \frac{1}{m!}\, \mathcal{L}_m \rho_{\,n-m}.
\end{equation}
Substituting Eq.~(\ref{3order}) into these relations, taking the trace, and integrating over time, we obtain
\begin{eqnarray}
J_i
&=& \lambda'
= \mathrm{Tr}\!\left( \mathcal{L}_1 \rho_s \right), \\
\mathrm{Var}(J_i)
&=& \lambda''
= \mathrm{Tr}\!\left( \mathcal{L}_2 \rho_s \right)
+ 2\, \mathrm{Tr}\!\left( \mathcal{L}_1 \varrho_{\,1} \right),
\end{eqnarray}
with
\begin{equation}
\varrho_{\,1}
= \mathcal{L}_0^{-1}
\big(
J_i \rho_s- \mathcal{L}_1 \rho_s
\big).
\end{equation}
Here $\mathcal{L}_0^{-1}$ denotes the generalized inverse of $\mathcal{L}_0$, restricted to the traceless subspace (equivalently, the subspace orthogonal to the steady state).

\subsection{Diagonal and Coherent Contributions to the Current Variance}

Decomposing the steady state into its diagonal and off-diagonal parts, one finds
\begin{eqnarray}\label{VdVc}
J_i
&=& \mathrm{Tr}\!\left( \mathcal{L}_1 \rho_d \right), \\
\mathrm{Var}(J_i)
&=&
\underbrace{
\mathrm{Tr}\!\left( \mathcal{L}_2 \rho_d \right)
+ 2\, \mathrm{Tr}\!\left[
\mathcal{L}_1 \mathcal{L}_0^{-1}
\left( J_i \rho_d - \mathcal{L}_1 \rho_d \right)
\right]}_{\equiv\,\mathrm{Var}_d(J_i)}
\;+\;
\underbrace{
2 J_i\, \mathrm{Tr}\!\left(
\mathcal{L}_1 \mathcal{L}_0^{-1}
\rho_c
\right)}_{\equiv\,\mathrm{Var}_c(J_i)} .
\end{eqnarray}
The current $J_i$ is entirely determined by the diagonal component,
whereas the variance $\mathrm{Var}(J_i)$ naturally separates into a diagonal
(classical) contribution and a coherent contribution originating from
$\rho_c$.
Importantly, the diagonal contributions
coincides exactly with the result obtained by applying the same counting
procedure to the classical master equation~(\ref{MasterEqClassical}).
This follows from the fact that $\mathcal{L}_1$ and $\mathcal{L}_2$ take the same
form in the classical counterpart and that $\rho_d$ coincides with its steady
state.
In addition, the term 
$\mathcal{L}_0^{-1} \left( J_i \rho_d - \mathcal{L}_1 \rho_d \right)
= J_i (\xi_d - \mathrm{Tr}\,\xi_d\,\rho_0)$,
where the diagonal operator $\xi_d$ satisfies
\begin{eqnarray} 
\mathcal{D}(\xi_d)
= \rho_d -   \mathcal{L}_1 \rho_d/{J_i}, \quad  \rm{and} \quad  \langle \Phi_0 | \xi_d | \Phi_0 \rangle
=
\langle \Phi_1 | \xi_d | \Phi_1 \rangle .
\end{eqnarray} 
The operator $\xi_d$ takes the same form in both the quantum model and its
classical counterpart, and consequently
$\mathrm{Tr}\!\left( \mathcal{L}_1 \xi_d \right)=   \mathrm{Tr} \mathcal{L}_1 (\xi_d - \mathrm{Tr}\,\xi_d\,\rho_0)=   \mathrm{Tr} (\mathcal{L}_1 \xi_d )-  J_i \mathrm{Tr}\,\xi_d  $
is identical in the two descriptions.

The coherent contribution to the current variance, $\mathrm{Var}_c(J_i)$,
can be obtained by evaluating $\mathcal{L}_0^{-1}\rho_c$.
To this end, we adopt the ansatz
\begin{equation}
\mathcal{L}_0^{-1}\rho_c
=
a\,(\tau-\rho_0)
+
b\left(
|\Phi_0\rangle\langle\Phi_1|
+
|\Phi_1\rangle\langle\Phi_0|
\right),
\end{equation}
which ensures that the action of $\mathcal{L}_0$ generates coherences only within
the virtual-qubit subspace and preserves the trace condition.
Acting with $\mathcal{L}_0$ then yields
\begin{equation}
-\gamma\, b = x, \qquad
- g r_0\, a + \Delta\, b = y ,
\end{equation}
where $r_0=P_0-P_1$.
Solving this linear system for $a$ and $b$, and substituting the expressions
for $x$ and $y$ given in Eq.~(\ref{xy}), one obtains
\begin{equation}
\mathrm{Var}_c(J_i)
= - 2a J^2_i
= -2\frac{1}{\gamma}
\frac{\gamma^2 - \Delta^2}{\gamma^2 + \Delta^2}
\frac{r}{r_0} J^2_i.
\end{equation}
 
The classical part, $\mathrm{Var}_d(J_i)$,
can be simplified within the classical model by using the current
associated with the dissipator $\mathcal{D}_c$.
Following the procedure outlined in Appendix~\ref{G0G12}, one can show that
the currents satisfy the relations
\begin{equation}
J_i = - n_i J_c , \qquad
\mathrm{Var}_d(J_i) = n_i^2\, \mathrm{Var}(J_c) .
\end{equation}
Using Eqs.~(\ref{VdVc}), we then obtain
\begin{equation}
J_c =  \frac{1}{2} p_c\, r ,
\end{equation}
and
\begin{equation}
\mathrm{Var}(J_c)
=
\frac{1}{2} p_c (q_0 + q_1)
- 2 J_c^2 \,\mathrm{Tr}\,\xi_d .
\end{equation}
We emphasize that the operator $\xi_d$ depends explicitly on the
superoperator $\mathcal{L}_1$, and therefore on the choice of the reservoir
under consideration.

\section{Optimization of $\mathcal{Q}_c$}\label{Qcmin}

In this part, we show how the coherent coupling $H_I$ should be chosen to minimize the
coherent contribution $\mathcal{Q}_c$ for fixed $H_0$ and dissipators $\{D_i\}$.
To this end, we analyze the structure of the steady state.

In the regime $p_c \gg p_i$  (equivalently, $g \to \infty$),
the diagonal part of the steady state $\rho_d$ approaches a well-defined limit
$\rho_I$, satisfying
\begin{eqnarray}\label{rhoI}
&&\mathcal{D}(\rho_I)
=
\frac{1}{2} p_I r_0
\left(
|\Phi_0\rangle\langle\Phi_0|
-
|\Phi_1\rangle\langle\Phi_1|
\right),\nonumber\\
&&\langle \Phi_0 | \rho_I | \Phi_0 \rangle
=
\langle \Phi_1 | \rho_I | \Phi_1 \rangle
=
\frac{d}{2},
\end{eqnarray}
where the parameters $p_I$ and $d$ are fully determined by the dissipative rates
$\{p_i, R_i\}$.
The diagonal part of the steady state, $\rho_d$, which also coincides with the steady state of
the corresponding classical master equation~(\ref{MasterEqClassical}), can then be written as
\begin{equation}\label{rhod}
\rho_d
=
\frac{p_I}{p_I + p_c}\, \tau
+
\frac{p_c}{p_I + p_c}\, \rho_I .
\end{equation}
As a consequence, one finds the relation
\begin{equation}
p_I r_0 = (p_I + p_c)\, r .
\end{equation}
This condition also implies that positivity of the density matrix requires $p_I>0$;
otherwise, $r$ would diverge as $p_I+p_c \to 0$.
The off-diagonal part of the steady state is determined by
Eqs.~(\ref{rhoc}) and~(\ref{xy}), with the coherence magnitude given by
\begin{equation}\label{Coh}
|\langle \Phi_0 | \rho_0 | \Phi_1 \rangle|^2
=
\frac{g^2}{\gamma^2 + \Delta^2} r^2
=
\frac{ 1}{4\gamma}\, p_c r^2 .
\end{equation}
In the limit $p_c \to \infty$, this quantity vanishes.
Therefore, $\rho_I$ also represents the limiting form of the full steady state $\rho_0$
in the strong-coupling regime.

%

The expressions for the coherence magnitude in Eq.~(\ref{Coh})  and for the
coherent contribution $\mathcal{Q}_c$ in Eq.~(\ref{QdQc}) show that maximizing the
steady-state coherence is a necessary condition for minimizing $\mathcal{Q}_c$.
Moreover, the maximal coherence is independent of the detuning $\Delta$.
This can be seen most clearly by taking $p_c$ (replacing $g$) and $\Delta$ as
independent control parameters.
For a fixed detuning $\Delta$, optimizing the coherent contribution $\mathcal{Q}_c$
is equivalent to maximizing the factor $p_c r^2$, which directly quantifies the
steady-state coherence.
Conversely, for a fixed $p_c$, the coherent contribution $\mathcal{Q}_c$ is minimized
at $\Delta = 0$.
Since the parameters $p_c$ and $r$ are fully correlated, it is convenient to further
replace $p_c$ by $r$ as the independent variable, which leads to
\begin{equation}
p_c r^2 = p_I r (r_0 - r).
\end{equation}
This expression attains its maximum at $r = r_0/2$, yielding
\begin{equation}
(p_c r^2)_{\max} = \frac{1}{4} p_I r_0^2 .
\end{equation}


\section{Optimal interaction strength minimizing $\mathcal{Q}$}\label{Qmin}

For fixed $H_0$, dissipators $\{D_i\}$, and detuning $\Delta$, we replace the coupling strength $g$
(or equivalently $p_c$) by the parameter $r$ as the independent control variable.
It can be shown that, similarly to the coherent contribution $\mathcal{Q}_c$, the diagonal
part $\mathcal{Q}_d$ is also a quadratic function of $r$.
As a result, the optimal interaction strength minimizing the total thermodynamic uncertainty
$\mathcal{Q}$ can be obtained analytically.
Below we outline the derivation.

The diagonal uncertainty $\mathcal{Q}_d$ can be expressed in terms of the current associated with
the dissipator $D_c$ as
\begin{equation}\label{Qd}
\mathcal{Q}_d
=
\ln \frac{P_0}{P_1}
\left(
\frac{\mathrm{Tr}\,\mathcal{L}_2 \rho_d}{\mathrm{Tr}\,\mathcal{L}_1 \rho_d}
-
2 J_c\, \mathrm{Tr}\,\xi_d
\right),
\end{equation}
where the  operator $\xi_d$ is defined through
\begin{eqnarray}\label{xid}
\mathcal{D}(\xi_d)
=
\rho_d - \mathcal{L}_1 \rho_d / J_c, \ \ \ \ \ 
\langle \Phi_0 | \xi_d | \Phi_0 \rangle
=
\langle \Phi_1 | \xi_d | \Phi_1 \rangle .
\end{eqnarray}
Here, the superoperators $\mathcal{L}_1$ and $\mathcal{L}_2$ are defined in Eq.~(\ref{Ln}) with the
index $i=c$.
The associated current is given by
\begin{equation}
J_c
=
\mathrm{Tr}\,\mathcal{L}_1 \rho_d
=
\frac{1}{2} p_c r
=
\frac{1}{2} p_I (r_0 - r).
\end{equation}
Acting on a diagonal state and taking the trace, the superoperator $\mathcal{L}_1$ extracts the
$\Sigma_z=
|\Phi_0\rangle\langle\Phi_0|
-
|\Phi_1\rangle\langle\Phi_1|$ 
component in the virtual qubit subspace,
and satisfies the relation $\mathcal{L}_1^2 \propto \mathcal{L}_2$.
As a result, one can decompose $\mathcal{L}_1 \rho_d$ as
\begin{equation}\label{L1rhod}
\mathcal{L}_1 \rho_d
=
\mathrm{Tr}(\mathcal{L}_1 \rho_d)\, \frac{1}{2}\Sigma_0
-
\mathrm{Tr}(\mathcal{L}_2 \rho_d)\, \frac{1}{2}\Sigma_z,
\end{equation}
where
$
\Sigma_0
=
|\Phi_0\rangle\langle\Phi_0|
+
|\Phi_1\rangle\langle\Phi_1|.
$
The generalized inverse $\tilde{\mathcal{D}}^{-1}$ is linearmap that sends a traceless diagonal
operator $X$ to a diagonal operator $Y$ satisfying $\mathcal{D}(Y)=X$ and 
$\langle \Phi_0 | Y | \Phi_0 \rangle
=
\langle \Phi_1 | Y | \Phi_1 \rangle .$
From Eq.~(\ref{rhoI}), one finds that
$ 
\tilde{\mathcal{D}}^{-1} \Sigma_z \propto \rho_I .
$ 
Combining the contribution proportional to $\Sigma_z$ in Eq.~(\ref{L1rhod}) with the first term in
Eq.~(\ref{Qd}), and regrouping  $\Sigma_0$  and 
$\rho_d$ in Eq.~(\ref{xid}), one obtains 
\begin{eqnarray}
\frac{\mathrm{Tr}\,\mathcal{L}_2 \rho_d}{\mathrm{Tr}\,\mathcal{L}_1 \rho_d}
-
2 J_c\, \mathrm{Tr}\,\xi_d
&=&
\frac{\mathrm{Tr}\,\mathcal{L}_2 \tau}{\mathrm{Tr}\,\mathcal{L}_1 \tau}
+
\frac{p_c}{p_I}\,
\frac{\mathrm{Tr}\,\mathcal{L}_2 \rho_I}{\mathrm{Tr}\,\mathcal{L}_1 \tau}
-\frac{2 \mathrm{Tr}\,\mathcal{L}_2 \rho_d}{p_I r_0}
-
2 J_c\,
\mathrm{Tr}\,
\tilde{\mathcal{D}}^{-1}
\!\left[
\frac{r}{r_0}
\left(
\tau - \frac{\Sigma_0}{2}
\right)
+
\frac{r_0-r}{r_0}
\left(
\rho_I - \frac{\Sigma_0}{2}
\right)
\right] \ \ \ \ \ 
\nonumber\\
&=&\frac{P}{r_0} +\frac{ r_0-r }{r_0 } (-r +\frac{d-P}{r_0})
-
p_I ( r_0-r) \mathrm{Tr}\,
\tilde{\mathcal{D}}^{-1}
\!\left[
\frac{r}{r_0}
\left(
\tau - \frac{\Sigma_0}{2}
\right)
+
\frac{r_0-r}{r_0}
\left(
\rho_I - \frac{\Sigma_0}{2}
\right)\right],
\end{eqnarray}
where $P=P_0+ P_1$ and $d/2=\langle \Phi_0 | \rho_I | \Phi_0 \rangle$.
Therefore, $\mathcal{Q}_d$ can always be written in the form
\begin{equation}\label{Qdr}
\mathcal{Q}_d
=
\ln \frac{P_0}{P_1} \frac{P}{r_0}
\left[
1
+
(r_0-r)\bigl(\mathcal{A} r + \mathcal{B}\bigr)
\right],
\end{equation}
where the coefficients $\mathcal{A}$ and $\mathcal{B}$ depend only on the
dissipators $\{D_i\}$ of the model.

To achieve the minimum of the thermodynamic uncertainty, we set the detuning $\Delta = 0$. In this case, the total uncertainty reduces to
\begin{equation}
\mathcal{Q}\label{Qr}
=
\ln \frac{P_0}{P_1}\,
\frac{P}{r_0}
\left[
1
+
(r_0 - r)\bigl(\mathcal{A}_1 r + \mathcal{B}\bigr)
\right],
\end{equation}
where 
$\mathcal{A}_1
=
\mathcal{A}
-
\frac{p_I}{P\,\gamma}.$
The extremum of $\mathcal{Q}$ is attained at
\begin{equation}
r_\ast
=
\frac{r_0}{2} -\frac{ \mathcal{B}}{2\mathcal{A}_1},
\end{equation}
with the corresponding extremal value given by
\begin{equation}
\mathcal{Q}(r_\ast)
=
\ln \frac{P_0}{P_1}\,
\frac{P}{r_0}
\left[
1
+
\frac{(\mathcal{A}_1 r_0 + \mathcal{B})^2}{4\mathcal{A}_1}
\right].
\end{equation}
For all examples considered, the coefficient $\mathcal{A}_1$ is always negative, and the
extremum $r_\ast$ lies within the interval $0 < r_\ast \le r_0/2$.
This range corresponds to $0 < p_c \le p_I$, or equivalently to
$4 g^2 \le \gamma p_I$.
\textit{We therefore conjecture that this behavior is generic.}
As a result, $\mathcal{Q}(r_\ast)$ gives the minimal uncertainty attainable for fixed
$H_0$ and dissipators $\{D_i\}$ by optimizing the coherent coupling $H_I$.

\section{Thermodynamic uncertainty in the vicinity of the Carnot limit}\label{r00}

At $r_0 = 0$, the coherent coupling does not modify the steady state of the system.
In this limit, one has
\begin{equation}
\rho_s = \tau = \rho_d = \rho_I, \qquad \rho_c = 0,
\end{equation}
and the associated current vanishes, $J_c = 0$.
Consequently, the coherent part of the thermodynamic uncertainty 
$\mathcal{Q}_c = 0$.
The diagonal contribution satisfies $\mathcal{Q}_d = 2$, which follows from linear
response theory for classical stochastic processes \cite{Barato2015Thermodynamic}.
This result can also be  derived directly within our formalism.
One can substitute $\tau = \lim_{r_0 \to 0} \rho_d$ into Eq.~(\ref{Qd}) and obtain
\begin{eqnarray}
\mathcal{Q}_d(r_0 = 0)
&=&
\lim_{r_0 \to 0}
\ln \frac{P_0}{P_1}
\left[
\frac{P}{r_0}\frac{p_I + p_c}{p_I}
-
\mathrm{Tr}\,
\tilde{\mathcal{D}}^{-1}
\!\left(
2 J_c \tau - 2 \mathcal{L}_1 \tau
\right)
\right]
\nonumber\\
&=&
\lim_{r_0 \to 0}
\ln \frac{P_0}{P_1}
\left[
\frac{P}{r_0}\frac{p_I + p_c}{p_I}
-
\mathrm{Tr}\,
\tilde{\mathcal{D}}^{-1}
\!\left(
\frac{1}{2} p_c P \Sigma_z
\right)
\right]
\nonumber\\
&=&
\lim_{r_0 \to 0}
\ln \frac{P_0}{P_1}
\left[
\frac{P}{r_0}\frac{p_I + p_c}{p_I}
-
\frac{P}{r_0}\frac{p_c}{p_I}
\right]
\nonumber\\
&=&
2 ,
\end{eqnarray}
where the transition from the second to the third line follows from
Eq.~(\ref{rhoI}).

The value of $r_0$ depends on $\{R_i\}$ associated with those
transition channels from $|\Psi_0\rangle$ to $|\Psi_1\rangle$ for which $n_i \neq 0$.
One may therefore choose $r_0$ as an independent control parameter by replacing one
of $R_i$, and perform a Taylor expansion of both $\mathcal{Q}_c$ and
$\mathcal{Q}_d$ around $r_0 = 0$.
In doing so, it is convenient to factor out the common prefactor
$\ln\!\left(P_0/P_1\right)\, P/r_0$, whose expansion up to second order in $r_0$ reads
\begin{equation}
\ln\!\left(\frac{P_0}{P_1}\right)\frac{P}{r_0}
\simeq
2\left[
1
+
\frac{1}{3}
\left(
\frac{r_0}{P}
\right)^2
\right].
\end{equation}
Using Eq.~(\ref{QdQc}), the coherent contribution $\mathcal{Q}_c$ can be written as
\begin{equation}
\mathcal{Q}_c
=
\ln \frac{P_0}{P_1}\,
\frac{P}{r_0}\,
\mathcal{F}_c\, r_0^2 ,
\end{equation}
where
\begin{equation}
\mathcal{F}_c
=
-
\left(
\frac{p_I}{p_I + p_c}
\right)^2
\frac{p_c}{P\,\gamma}
\frac{\gamma^2 - \Delta^2}{\gamma^2 + \Delta^2}.
\end{equation}
The value of $\mathcal{F}_c$ evaluated at $r_0 = 0$ therefore gives the coefficient of
the quadratic term in the expansion of $\mathcal{F}_c r_0^2$.
Moreover, in Eq.~(\ref{Qdr}), the leading contribution of the factor
$(r_0 - r)\bigl(\mathcal{A}  r + \mathcal{B}\bigr)$ is necessarily of second order
in $r_0$, and can be expressed in the vicinity of $r_0 = 0$ as
\begin{equation}
(r_0 - r)\bigl(\mathcal{A}  r + \mathcal{B}\bigr)
\simeq
\mathcal{F}_d\, r_0^2 ,
\end{equation}
since $r_0 = 0$ corresponds to an extremal point of $\mathcal{Q}_d$.
Collecting the above results, the thermodynamic uncertainty expanded up to second
order in $r_0$ takes the form
\begin{equation}
\mathcal{Q}
\simeq
2\left[
1
+
\frac{1}{3}
\left(
\frac{r_0}{P}
\right)^2
+
\mathcal{F}_c\, r_0^2
+
\mathcal{F}_d\, r_0^2
\right].
\end{equation}
We remark that the quantities $P$, $\mathcal{F}_c$, and $\mathcal{F}_d$ are
independent of the particular choice of free control parameters, and are fully
determined by the system Hamiltonian and interaction parameters evaluated at
$r_0 = 0$.

We summarize the above analysis as follows.
In the vicinity of the Carnot limit, the violation of the classical bound
$\mathcal{Q} \geq 2$ occurs if and only if the quadratic coefficient in the expansion
of $\mathcal{Q}$ becomes negative, namely,
\begin{equation}
\frac{1}{3 P^2}
+
\mathcal{F}_c
+
\mathcal{F}_d
< 0 .
\end{equation}
At the optimal interaction strength, the combined coefficient satisfies
\begin{equation}
\mathcal{F}_c + \mathcal{F}_d
=
\lim_{r_0 \to 0}
\frac{(\mathcal{A}_1 r_0 + \mathcal{B})^2}{4 \mathcal{A}_1 r_0^2} .
\end{equation}
Therefore, for a fixed connectivity structure between the system eigenstates and the
reservoirs, a sufficient condition for achieving $\mathcal{Q} < 2$ is the existence of
a parameter regime in which
\begin{equation}
\lim_{r_0 \to 0}
\frac{(\mathcal{A}_1 r_0 + \mathcal{B})^2}{4 \mathcal{A}_1 r_0^2}+\frac{1}{3 P^2}
<0.
\end{equation}
is satisfied.

\section{Additional physical examples}
\label{Examples}


\subsection{Driven qutrit as a quantum heat engine}

 \begin{figure}
 \includegraphics[width=4.56cm]{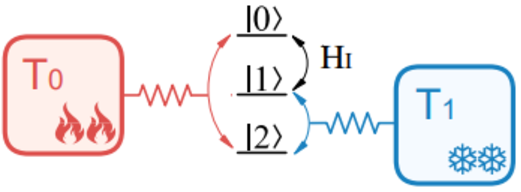}
\caption{
(Color online)
Schematic illustration of a qutrit (three-level system) operating as a quantum heat engine between two thermal reservoirs.
The two levels $|0\rangle$ and $|2\rangle$ are coupled to a thermal bath at temperature $T_0$, while $|1\rangle$ and $|2\rangle$ are coupled to another bath at temperature $T_1$.
In this model, the virtual qubit is formed by the two states
$|\Phi_0\rangle = |0\rangle$ and $|\Phi_1\rangle = |1\rangle$,
which are coherently driven by $H_I$ [Eq.~(\ref{HIt})].
For $T_0 > T_1$, and within appropriate parameter regimes, the system can operate as a heat engine that delivers work to the driving field; alternatively, it can function as a refrigerator, where work is performed by the driving field on the qutrit to extract heat from the cold reservoir.
}
 \label{FigModel3_1qutrit}
\end{figure}

As a third example, we consider a three-level system (qutrit) weakly coupled to two
thermal reservoirs and subject to a coherent driving between its two states \cite{Kalaee2021},
as schematically shown in Fig.~\ref{FigModel3_1qutrit}.
The bare Hamiltonian of the qutrit reads
\begin{equation}\label{H3}
H_0 = \sum_{i=0}^{2} E_i \, |i\rangle\langle i| ,
\end{equation}
where $|i\rangle$ ($i=0,1,2$) denote the three energy eigenstates.
The virtual qubit in this model is
formed by the two states $|\Phi_0\rangle = |0\rangle$ and $|\Phi_1\rangle = |1\rangle$,
which are coherently driven by $H_I$ [Eq.~(\ref{HIt})]. 
The qutrit is coupled to two thermal reservoirs.
Each pair of states $\{|i\rangle, |2\rangle\}$ ($i=0,1$) interacts with a reservoir at temperature $T_i$,
exchanging photons of frequency $\omega_i = E_i - E_2$.
No specific ordering of   $E_i$ is assumed here;
when $\omega_i < 0$, the corresponding process should be understood as a transition
from $|i\rangle$ to $|2\rangle$ accompanied by the absorption of a photon of energy $|\omega_i|$.
The dissipators associated with the two reservoirs take the form of Eq.~(\ref{Di}), with jump operators $\Gamma_i = |2\rangle\langle i|$, coupling strengths $p_i$, and thermal occupation factors $R_i = 1/(1+\exp(\omega_i/T_i))$ and $\bar R_i = 1-R_i$. The corresponding decoherence rate of the virtual qubit is $\gamma = \frac{1}{2} (p_0 \bar R_0 + p_1 \bar R_1)$.
Transitions between the virtual qubit states can occur via the excited state $|2\rangle$, specifically through the sequence $|0\rangle \to |2\rangle \to |1\rangle$, with $n_0 = \omega_0/|\omega_0|$ and $n_1 = -\omega_1/|\omega_1|$.

Following the general procedure outlined in the main text, the steady states in the two limiting cases are given by
\begin{equation}
\tau = P_0 |0\rangle\langle 0| + P_1 |1\rangle\langle 1| + P_2 |2\rangle\langle 2|, \qquad
\rho_I = \frac{d}{2} \left( |0\rangle\langle 0| + |1\rangle\langle 1| \right) + (1-d) |2\rangle\langle 2|,
\end{equation}
with
\begin{equation}
P_0 = \frac{R_0 \bar R_1}{1-R_0 R_1}, \quad
P_1 = \frac{\bar R_0 R_1}{1-R_0 R_1}, \quad
P_2 = \frac{\bar R_0 \bar R_1}{1-R_0 R_1}, \qquad
d =2 \frac{p_0 R_0 + p_1 R_1}{p_0 + p_1 + p_0 R_0 + p_1 R_1},
\end{equation}
and
\begin{equation}
p_I = \frac{2 p_0 p_1 (1 - R_0 R_1)}{p_0 + p_1 + p_0 R_0 + p_1 R_1}.
\end{equation}
By forming the convex combination of $\tau$ and $\rho_I$ to obtain $\rho_d$, we can then compute the operator $\xi_d$ defined in Eq.~(\ref{xidzw}) (also Eq.~(\ref{xid}) in the Appendix).  
It can be expressed in the form
\begin{equation}
\xi_d = x_I \, \rho_I + x_2 \, |2\rangle\langle 2|,
\end{equation}
where
\begin{equation}
x_I = \frac{1}{P_I \, r_0} (\frac{q}{r}+r)+\frac{q_2}{P_I \, r_0}\frac{p_0 \, R_0-p_1 \, R_1}{p_0 \, R_0+p_1 \, R_1}, \qquad x_2 = \frac{-q_2}{p_0 \, R_0+p_1 \, R_1}.
\end{equation}
 Substituting these results into Eq.~(\ref{QdQc}), one can evaluate the two contributions
to the thermodynamic uncertainty.
The diagonal contribution $\mathcal{Q}_d$ can be further rearranged into the form
given in Eq.~(\ref{Qdr}),
with the coefficients
\begin{equation}
\mathcal{A}
=
-
2
\frac{
p_0^2(1+R_0)+p_1^2(1+R_1)+p_0 p_1 (R_0+R_1+2R_0R_1)
}{
\bigl[p_0(1+R_0)+p_1(1+R_1)\bigr]^2
},
\quad
\mathcal{B}
=
-
4
\frac{
p_0 p_1 (R_0-R_1)
}{
\bigl[p_0(1+R_0)+p_1(1+R_1)\bigr]^2
}.
\end{equation}

The minimum value of the thermodynamic uncertainty for this model is found numerically
to be $\mathcal{Q}_{\min} = 1.549$,
which is achieved at
$\Delta = 0$, $p_1/p_0 = 0.83$, $R_0 = 0.946$, $R_1 = 0.129$, and $r/r_0 = 0.421$.
For these parameters, the thermal occupations correspond to the energy ordering
$E_0 > E_2 > E_1$.
In this regime, however, the system cannot operate as either a heat engine or a refrigerator.
If the model is required to operate as a heat engine or a refrigerator,
the energy levels $E_0$ and $E_1$ must lie on the same side of $E_2$.
Under this constraint, the minimum thermodynamic uncertainty increases slightly and is found numerically
to be $\mathcal{Q}_{\min} = 1.618$.
This minimum is attained at
$\Delta = 0$, $p_1/p_0 = 0.33$, $R_0 = 0.979$, $R_1 \to 0.5^{+}$ (corresponding to $T_1 \to \infty$),
and $r/r_0 = 0.477$.

In the reversible limit, $r_0 \to 0$, the thermodynamic uncertainty can be expanded as
\begin{equation}
\mathcal{Q} \simeq 2 + \delta\, r_0^2 ,
\end{equation}
where the coefficient $\delta$ attains its minimum at $\Delta = 0$ and $p_0 = p_1$,
\begin{equation}
\delta_{\min}
=
\frac{4 - 15 R - 6 R^2 + R^3}{24 R^2 (1 + R)} ,
\end{equation}
with $R = R_0 = R_1$.
The coefficient $\delta_{\min}$ becomes negative for $4 - 15 R - 6 R^2 + R^3 < 0$,
namely for $R > 0.244$.
This provides a sufficient condition under which the model can surpass the classical bound.

\subsection{Driven qutrit in the global master-equation approach}\label{global}

Despite the fact that Ref.~\cite{bayona2021thermodynamics} adopts a master equation derived
within the global approach, the \emph{resonant-coupling case} identified there—under which
the TUR can be violated—is equivalent to the model
discussed above. 
As a result, the corresponding behavior can be captured within the same
formal framework developed in this work.
While this equivalence is demonstrated explicitly for the driven qutrit model, the underlying reasoning naturally extends to the broader class of driven virtual-qubit models considered in the present work.

Specifically, the model illustrated in Fig.~\ref{FigModel3_1qutrit} satisfies, within
the global approach of Ref.~\cite{bayona2021thermodynamics}, the master equation
\begin{equation}\label{Meq3}
\dot{\rho}
=
-i[H(t),\rho]
+
\mathcal{D}(\rho),
\end{equation}
where $H(t)=H_0+H_I$.
Here, $H_0$ is the bare system Hamiltonian given in Eq.~(\ref{H3}), and the coherent driving term reads
\begin{equation}
H_I
=
g \left(
|0\rangle\langle 1| e^{-i\omega_d t}
+
|1\rangle\langle 0| e^{i\omega_d t}
\right).
\end{equation}
The dissipator $\mathcal{D}$ consists of four contributions (two for each thermal reservoir),
each of which takes the Lindblad form given in Eq.~(\ref{Di}). The corresponding jump operators
describe transitions between the instantaneous eigenstates
$\{|\epsilon_i(t)\rangle\}$ of the driven Hamiltonian $H(t)$.
The resonant-coupling case considered in Ref.~\cite{bayona2021thermodynamics} corresponds to the
situation in which, for the $i$th reservoir ($i=0,1$), only a single dissipative channel is
active. In this case, the jump operator can be written as
\begin{equation}
\Gamma_i
=
|\epsilon_2(t)\rangle\langle \epsilon_i(t)|,
\end{equation}
indicating that the reservoir induces transitions exclusively between the instantaneous
eigenstates $|\epsilon_i(t)\rangle$ and $|\epsilon_2(t)\rangle$.

One can construct a unitary transformation
$\mathcal{R}_d = \mathcal{R}_0 \mathcal{R}_s$
that diagonalizes the time-dependent Hamiltonian,
\begin{equation}
\mathcal{R}_d H(t) \mathcal{R}_d^{\dagger}
=
\tilde{H}
=
\sum_{i=0}^{2} \epsilon_i \, | i \rangle \langle i | .
\end{equation}
Here, $\mathcal{R}_s$ denotes the time-dependent rotation operator introduced
in Eq.~(\ref{Rszw}), with generator
\begin{equation}
G = -\frac{\omega_d}{2}
\left(
|0\rangle\langle 0| - |1\rangle\langle 1|
\right),
\end{equation}
while $\mathcal{R}_0$ is a static rotation acting within the
$\{|0\rangle, |1\rangle\}$ subspace,
\begin{equation}
\mathcal{R}_0
=
\exp\!\left[
\theta \left(
 |0\rangle\langle 1| - |1\rangle\langle 0|
\right)
\right] +|2 \rangle\langle 2|,
\qquad
\tan(2\theta) = \frac{2g}{E_0 - E_1}.
\end{equation}

The corresponding instantaneous eigenvalues read
\begin{equation}
\epsilon_{0,1}
=
\frac{E_0 + E_1}{2}
\pm
\frac{1}{2}
\sqrt{(E_0 - E_1)^2 + 4 g^2},
\qquad
\epsilon_2 = E_2,
\end{equation}
and 
the instantaneous eigenstates are given by
$|\epsilon_i(t)\rangle = \mathcal{R}_d^{\dagger} | i \rangle$,
which explicitly read
\begin{equation}
|\epsilon_0(t)\rangle
 =
\cos\theta\, e^{-i\frac{\omega_d }{2}t} |0\rangle
+
\sin\theta\,e^{ i\frac{\omega_d }{2}t} |1\rangle,  \qquad
|\epsilon_1(t)\rangle
 =
-\sin\theta\, e^{-i\frac{\omega_d }{2}t} |0\rangle
+
\cos\theta\, e^{i\frac{\omega_d }{2}t} |1\rangle, \qquad
|\epsilon_2(t)\rangle
 =
|2\rangle .
\end{equation}

In the instantaneous eigenbasis, the density matrix is defined as
$\tilde{\rho} = \mathcal{R}_d \rho \mathcal{R}_d^{\dagger}$.
Substituting this expression into Eq.~(\ref{Meq3}), we obtain the master equation
in the instantaneous eigenstate representation,
\begin{equation}
\dot{\tilde{\rho}}
=
-i[\tilde{H}, \tilde{\rho}]
-i[\tilde{G}, \tilde{\rho}]
+
\tilde{\mathcal{D}}(\tilde{\rho}) .
\end{equation}
Here, the dissipator $\tilde{\mathcal{D}}$ is obtained from
$\mathcal{D}$ in Eq.~(\ref{Meq3}) by replacing the jump operators
$|\epsilon_j(t)\rangle\langle \epsilon_i(t)|$ with $|j\rangle\langle i|$
and
\begin{equation}
\tilde{G} = \mathcal{R}_0 G \mathcal{R}_0^{\dagger} = -\frac{1}{2}\omega_d \cos 2\theta  \left(
|0\rangle\langle 0| - |1\rangle\langle 1|
\right) + \frac{1}{2}\omega_d \sin 2\theta  \left(
|0\rangle\langle 1| + |1\rangle\langle 0|
\right) 
\end{equation}
%
Under the resonant-coupling condition defined in Ref.~\cite{bayona2021thermodynamics}, where the $i$th reservoir induces transitions
exclusively between the instantaneous eigenstates
$|\epsilon_i(t)\rangle$ and $|\epsilon_2(t)\rangle$,
the above master equation takes the form of Eq.~(\ref{MasterEqRot}).
It is therefore equivalent to the driven qutrit model analyzed in the previous section
within the local master-equation approach,
with the bare system Hamiltonian given by $\tilde{H}$,
a coherent coupling strength $\tilde{g}=\frac{1}{2}\omega_d \sin 2\theta $,
and a detuning parameter $\tilde\Delta = \epsilon_0 - \epsilon_1 - \omega_d \cos 2\theta$.

 \subsection{Three-qubit absorption refrigerator }

 \begin{figure}
 \includegraphics[width=6.48cm]{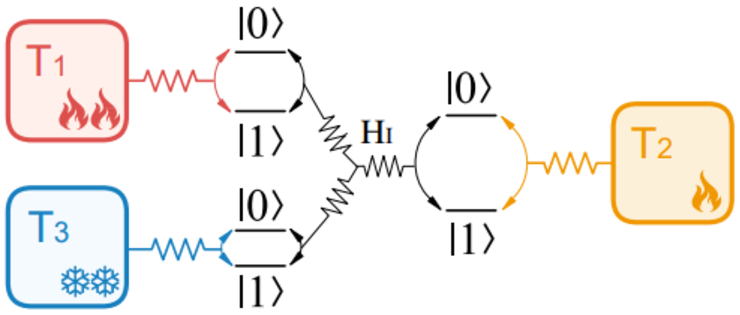}
\caption{
(Color online)
 Schematic illustration of the three-qubit absorption refrigerator .
Each qubit is weakly coupled to its own thermal reservoir at temperatures 
$T_i$, $i=1,2,3$.
In this model, the virtual qubit is formed by the two degenerate energy levels of the composite system,
namely $|\Phi_0\rangle = |010\rangle$ and $|\Phi_1\rangle = |101\rangle$.
They are coupled through a time-independent interaction of the form given in Eq.~(\ref{HIt}),
corresponding to $\omega_d = 0$.
 } \label{FigModel4_3qubit}
 \end{figure}

As a final example, we consider the three-qubit quantum absorption refrigerator \cite{PRL2010small,JPA2011smallest},
schematically illustrated in Fig.~\ref{FigModel4_3qubit}.
This model was originally proposed in the context of fundamental limits on the size
of thermal machines and has since motivated a large body of work on self-contained
quantum thermal devices.
The system consists of three qubits, each weakly coupled to an independent thermal
reservoir at temperatures $T_i$ $(i=1,2,3)$.
The purpose of the device is to extract heat from the cold reservoir attached to
qubit~$3$.

The bare system Hamiltonian is given by
\begin{equation}
H_0
=
\frac{1}{2}
\sum_{\alpha=1}^{3}
\omega_\alpha\, \sigma_\alpha^{z},
\end{equation}
where the transition frequencies satisfy the resonance condition
$\omega_1 + \omega_3 = \omega_2$.
In this model, the virtual qubit is formed by two degenerate energy eigenstates,
\(
|\Phi_0\rangle = |010\rangle
\)
and
\(
|\Phi_1\rangle = |101\rangle
\).
These states are coupled via the interaction Hamiltonian in Eq.~(\ref{HIt}),
with $\omega_d = 0$, such that the detuning is fixed to $\Delta = 0$.
The coupling to the reservoirs is described by three dissipators of the form given in
Eq.~(\ref{Di}), with coupling strengths $p_i$ and thermal occupation factors
\(
R_i = [1+\exp(\omega/T_i)]^{-1}.
\)
The corresponding jump operators are $\Gamma_i = \sigma_i^{-}$, which may equivalently be
expressed in the energy eigenbasis of the bare Hamiltonian.
The resulting decoherence rate of the virtual qubit is
\(
\gamma = (p_1 + p_2 + p_3)/2.
\)
A transition from $|010\rangle$ to $|101\rangle$ is accompanied by net excitations exchanged
with the three reservoirs,
\(
n_1 = 1,\;
n_2 = -1,\;
n_3 = 1.
\)

In the absence of the interaction $H_I$, each qubit thermalizes with its own reservoir, and
the steady state is given by
\begin{equation}
\tau = \tau_1 \otimes \tau_2 \otimes \tau_3,
\end{equation}
where $\tau_i$ ($i=1,2,3$) denotes the local thermal state in Eq.~(\ref{tau}), with excited-state
population $R_i$.
For finite interaction, the steady state, as well as its limiting form $\rho_I$,
can be obtained analytically. The corresponding parameter $p_I$ reads
\begin{equation}
p_I
=
\frac{
2\gamma \, p_1 p_2 p_3 \, (p_1 + p_2)(p_1 + p_3)(p_2 + p_3)
}{
\sum_{i\neq j}
w_{ij} \, p_i^2 p_j^2 (p_i + p_j)\bigl( 4\gamma - p_i - p_j \bigr)
+
p_1 p_2 p_3 \bigl( 8\gamma^3 + p_1 p_2 p_3 \bigr)
}.
\end{equation}
Here $w_{ij} = 1 + s_i s_j$, with $s_{1,3} = R_{1,3} - \bar R_{1,3}$ and
$s_{2} =- R_{2} + \bar R_{2}$.  
Following the general procedure outlined in the main text,
the thermodynamic uncertainty is then evaluated and cast into the form of
Eq.~(\ref{Qr}).
For simplicity, we list here the two coefficients for the symmetric case
$p_1 = p_2 = p_3$,
\begin{equation}
\mathcal{A}_1
=
-\frac{6(27+20P)}{(9+4P)^2 P},
\qquad
\mathcal{B}
=
-\frac{108 r_0}{(9+4P)^2 P}.
\end{equation}
For this model, the thermodynamic uncertainty does not violate the classical bound;
its minimum value is $\mathcal{Q}_{\min}=2$, attained in the reversible limit.
Expanding $\mathcal{Q} = 2 + \delta r_0^2$ around this limit,
the minimal value of the second-order coefficient $\delta$ is achieved at
$p_2 = p_3 \simeq 0.26\, p_1$ and $R_i \to 1/2$,
yielding $\delta_{\min} = 1.21$.

\end{document}